\documentclass[twoside,12pt]{article}
\usepackage{epsfig}

\def \HM {HEIDEL\-BERG-MOSCOW~}

\def \ndbd {neutrinoless double beta decay}

\newcommand{\hdmo}{HEIDELBERG-MOSCOW experiment~}
\def \etal {et al.}

\def \znbb {$0\nu\beta\beta$~}

\newcommand{\ba}[1]{\begin{eqnarray} \label{(#1)}}
\newcommand{\ea}{\end{eqnarray}}

\begin{document} 

\title{Search For Neutrinoless Double Beta Decay  
\protect\newline With Enriched $^{76}{Ge}$ 1990-2003 -- \HM-Experiment}

\author{H.V. Klapdor-Kleingrothaus$^{\mbox{*}}$$^{\mbox{a}}$, 
	I.V. Krivosheina$^{\mbox{a,b}}$, A. Dietz$^{\mbox{a}}$,\\ 
C. Tomei$^{\mbox{a,c}}$, O. Chkvoretz$^{\mbox{a}}$, H. Strecker$^{\mbox{a}}$\\}

\date{ }
\maketitle
\begin{center}
\begin{tabular}{c}                    
$^{\mbox{a}}$                         
Max-Planck Institut f\"ur Kernphysik, Heidelberg, Germany\\ 
$^{\mbox{b}}$                         
Institute of Radiophysical Research, Nishnij Novgorod, Russia\\ 
$^{\mbox{c}}$ 
University of L'Aquila, Italy\\
$^{\mbox{*}}$ 
Spokesman of the Collaboration;      
E-mail: klapdor@gustav.mpi-hd.mpg,\\
 Home-page: http://www.mpi-hd.mpg.de.non\_acc/
\end{tabular}                         
\end{center}

\begin{abstract}
	The \HM experiment, which is the most sensitive 
	double beta decay experiment since ten years 
	has been regularly continued until end of November 2003. 
	An analysis of the data has been performed 
	already until May 20, 2003. The experiment yields now, 
	on a 4$\sigma$ level, evidence for lepton number violation 
	and proves that the neutrino is a Majorana particle.
	It further shows that neutrino masses are degenerate. 
	In addition it puts several stringent constraints on other physics 
	beyond the Standard Model. 
	Among others it opens the door to test various 
	supersymmetric theory scenarios, for example
	it gives the sharpest limit on the parameter ${\lambda}'_{111}$ 
	in the R-parity violating part of the superpotential, 
	and gives information on the splitting 
	of the sneutrino-antisneutrino system.
	The result from the \HM experiment is consistent  
	with recent results from CMB investigations, 
	with high energy cosmic rays, with the result from the g-2 experiment 
	 and with recent theoretical work.
	It is indirectly supported by the analysis of other 
	Ge double beta experiments. 
	Recent criticism of various kind has been shown to be wrong, 
	among others by measurements performed in 2003 with a $^{214}{Bi}$ 
	source ($^{226}{Ra}$), by simulation of the background 
	in the range of Q$_{\beta\beta}$ by GEANT4, 
	and by deeper investigation of statistical features 
	such as sensitivity of peak search, and relevance 
	of width of window of analysis. 
\end{abstract}


\section{Introduction}

	Double beta decay is the most sensitive probe to test 
	lepton number conservation.
	Further it seems to be the only way to decide about 
	the Dirac or Majorana nature of the neutrino.
	
	Double beta decay can contribute decisively to the field 
	of neutrino physics also by setting an absolute scale 
	to neutrino masses, which cannot be observed  
	from neutrino oscillation experiments.

	The observable of double beta decay is the effective neutrino mass

\centerline{$\langle m \rangle = |\sum U^2_{ei}m^{}_i| = |m^{(1)}_{ee}| 
		      + e^{i\phi_2} |m^{(2)}_{ee}| 
		      + e^{i\phi_3} |m^{(3)}_{ee}|,
$}

\noindent
	  with $U^{}_{ei}$ 
	  denoting elements of the neutrino mixing matrix, 
	  $m_i$ neutrino mass eigenstates, and $\phi_i$  relative Majorana 
	  CP phases. It can be written in terms of oscillation parameters 
\cite{KKPS} 
\begin{eqnarray}
\label{1}
|m^{(1)}_{ee}| &=& |U^{}_{e1}|^2 m^{}_1,\\
\label{2}
|m^{(2)}_{ee}| &=& |U^{}_{e2}|^2 \sqrt{\Delta m^2_{21} + m^{2}_1},\\
\label{3}
|m^{(3)}_{ee}| &=& |U^{}_{e3}|^2 \sqrt{\Delta m^2_{32} 
				 + \Delta m^2_{21} + m^{2}_1}.
\end{eqnarray}

	The effective mass $\langle m \rangle$ is related with the 
	half-life for $0\nu\beta\beta$ decay via 
$\left(T^{0\nu}_{1/2}\right)^{-1}\sim \langle m_\nu \rangle^2$, 
        and for the limit on  $T^{0\nu}_{1/2}$
	deducible in an experiment we have 
\begin{eqnarray}
\label{4}	
T^{0\nu}_{1/2} \sim \epsilon \times a \sqrt{\frac{Mt}{\Delta E B}}, 
\end{eqnarray}

	Here $a$ is the isotopical abundance of the $\beta\beta$ emitter;
	$M$ is the active detector mass; 
	$t$ is the measuring time; 
	$\Delta E$ is the energy resolution; 
	$B$ is the background count rate 
	and $\epsilon$ is the efficiency for detecting a $\beta\beta$ signal. 
	Determination of the effective mass fixes the absolute 
	scale of the neutrino mass spectrum  
\cite{KKPS,KK60Y}. 

	The \HM experiment has been regularly continued in 2003. 
	It had to be stopped, 
	on November 30, 2003, according to contract. 
	Unfortunately the Kurchatov institute did not agree 
	to prolong the contract. 
	The experiment is already since 2001 operated only 
	by the Heidelberg group, which also performed the analysis 
	of the experiment from its very beginning.

	The experiment is {\it since ten years now} the most sensitive 
	double beta experiment worldwide. 
	In this report we will describe in section II 
	the evidence for \ndbd
	~~($0\nu\beta\beta$), found by an analysis of the \HM experiment 
	including the three more years of data taking.

	The result derived from the full data taken until May 20, 2003 is 
\begin{eqnarray}
\label{5}
	{\rm T}_{1/2}^{0\nu} = (0.69 - 4.18) \times 10^{25}
	{\rm y}~~ (99.73\% c.l.)
\end{eqnarray}
	with best value of 
	${\rm T}_{1/2}^{0\nu} = 1.19 \times 10^{25}~ y$.
	Thus double beta decay is the slowest nuclear decay process 
	observed until now in nature.
	Assuming the neutrino mass mechanism to dominate the decay 
	amplitude, we deduce

\vspace{-0.8cm}
\begin{eqnarray}
\label{6}
	\langle m_{\nu} \rangle = (0.24 - 0.58)\,eV~~ (99.73\% c.l.),
\end{eqnarray}
	with best value of 0.44\,eV. 
	This value we obtained using the nuclear matrix element of 
\cite{Sta90}.  
	Allowing for an uncertainty of $\pm 50\%$ of the matrix 
	elements (see 
\cite{KK02-Found,KK60Y}),
	this range widens to 

\vspace{-0.7cm}
\begin{eqnarray}
\label{7}
	\langle m_{\nu} \rangle = (0.1 - 0.9)\,eV
\end{eqnarray}

	The result (2) and (3) determines the neutrino mass 
	scenario to be degenerate 
\cite{KK-Sark01,KK-S03-WMAP}.
	The common mass eigenvalue follows then to be 
	$m_{com}= (0.14 - 3.6)\, eV~~ (99.73\%)$.

	The new results with three more years of statistics confirm 
	our earlier results 
\cite{KK02,KK02-PN,KK-antw02,KK-BigArt02,KK02-Found,Support-KK03-PL}  
	on a higher confidence level. 
	The signal is now seen on a 4.2$\sigma$ level (see section 2).

	If we allow for other mechanisms (see 
\cite{KK-InJModPh98,KK-SprTracts00,KK60Y,KKS-INSA02}), 
	the value given in eq. 
(\ref{6}),(\ref{7}) 
	has to be considered as an upper limit. 
	In that case very stringent limits arise for some other fields 
	of beyond standard model physics. To give an example, 
	it gives the sharpest limit on the Yukawa coupling 
	${\lambda}'_{111}$ in the R-parity violating part 
	of the superpotential 
\cite{HK-KK95-99}.
	It also gives information on R-parity conserving supersymmetry. 
	New R-parity conserving SUSY contributions to \znbb decay 
	occur at the level of box diagrams 
\cite{HK-KK97-98}.
	Double beta decay then yields information on 
	the mass splitting in the sneutrino-antisneutrino system 
\cite{HK-KK97-98}. 
	These constraints leave room for accelerator searches 
	for certain manifestations of the second and 
	third generation (B-L)-violating 
	sneutrino mass term, 
	but are most probably too tight for first generation 
	(B-L)-violating sneutrino masses to be searched for directly.
	It has been discussed recently 
\cite{Uehara02}
	that ~$0\nu\beta\beta$ decay by R-parity violating SUSY experimentally 
	may not be excluded, although this would require making R-parity 
	violating couplings generation dependent.

	We show, in section III that indirect support for the observed 
	evidence for neutrinoless double beta decay  
	evidence comes from analysis of other Ge double beta experiments 
	(though they are by far less sensitive, they yield independent 
	information on the background in the region of the expected signal).

\vspace{-0.5cm}
\begin{table}[ht]
\caption{Recent support of the neutrino mass deduced from 
	~$0\nu\beta\beta$ decay 
\protect\cite{KK02,KK02-PN,KK02-Found,Bi-KK03-NIM,Backgr-KK03-NIM}
	by other experiments, and by theoretical work.}
{\normalsize
\newcommand{\m}{\hphantom{$-$}}
\renewcommand{\arraystretch}{1.}
\setlength\tabcolsep{5.7pt}
\begin{center}
\begin{tabular}{|c|c|c|}
\hline
Experiment	& References	&	$m_{\nu}$ (degenerate $\nu$'s)(eV)	\\
\hline
$0\nu\beta\beta$ 	
&	\cite{KK02,KK02-PN,KK02-Found,Bi-KK03-NIM,Backgr-KK03-NIM}	
& 0.12 - 0.9\\
WMAP 	&	\cite{WMAP03,Hannes03}	&	$<$ 0.23, or 0.33, or 0.50\\
CMB &	\cite{CMB02}	&		$<$ 0.7 \\
CMB+LSS+X-ray gal. Clust.
&	\cite{Allen03-Wmap}	& $\sim$ 0.2 eV\\
SDSS + WMAP	&	\cite{Barger03}	& $<$ 0.57\,eV\\
Z - burst 
&	\cite{Farj00-04keV,FKR01}	& 0.08 - 1.3\\
g-2 &	\cite{MaRaid01}	&	$>$ 0.2 \\
Tritium &	\cite{Trit03}
		& $<$ 2.2 - 2.8\\
$\nu$ oscillation	
& \cite{KAML02,Fogli03}			&	$>$ 0.04\\
\cline{1-3}
	{\bf Theory:}	&	&	\\
\cline{1-3}
A$_4$-symmetry  &	\cite{BMV02}		&	$>$ 0.2 \\
identical quark &&\\
and $\nu$ mixing at GUT scale 
&	\cite{Moh03}	
&	$>$ 0.1 \\
Alternative cosmological 	& 				&	\\
'concordance model'	&	\cite{S-Sarkar03}	&	order of eV\\
\hline
\end{tabular}
\end{center}
\label{Confirm-exp}}
\end{table}


	The discussion in section IV, V, VI, may now just still 
	be of historical interest. Here we disprove some criticism 
	of our {\it earlier} given results. We show by measurements 
	with a $^{226}{Ra}$ source, performed in 2003 
\cite{Bi-KK03-NIM}, 
	and by various statistical calculations, 
	that the criticism 
	by Aalseth et al., (see Mod. Phys. Lett. A17 (2002) 1475-1478), 
	Zdesenko et al., (see Phys. Lett. B 546 (2002) 206-215), 
	Ianni (in NIM 2004), 
	Feruglio et al., (see Nucl. Phys. B 637 (2002) 345) 
	of our earlier results 
\cite{KK02,KK02-PN,KK02-Found}
	just was {\it wrong}.

	In section VII we give a short discussion, stressing that 
	the evidence for neutrinoless double beta decay 
	has been supported by various recent experimental results 
	from other fields of research 
	(see Table 
\ref{Confirm-exp}). 
	It is consistent 
\cite{KK-S03-WMAP}
	with recent results from cosmic 
	microwave background experiments 
\cite{CMB02,WMAP03,Hannes03}.
	The precision of WMAP~~ even allows to rule out some 
	old-fashioned nuclear double beta decay matrix elements (see 
\cite{Muray03}).

	It has been shown to be consistent with the neutrino masses 
	required for the Z-burst scenarios of high-energy cosmic rays 
\cite{FKR01,Farj00-04keV
}.
	It is consistent with a (g-2) deviating from the standard 
	model expectation 
\cite{MaRaid01}.
	It is consistent also with the limit from the tritium 
	decay experiments 
\cite{Weinh-Neu00
}
	but the allowed confidence range still 
	extends down to a range which cannot be  
	covered by future tritium experiments, 
	if at all  
\cite{Kirch04}.
	It is further supported by recent theoretical work 
\cite{BMV02,Moh03,Hof04}.

	Cosmological experiments like WMAP are now on the level 
	that they can seriously contribute to terrestrial research. 
	The fact that WMAP and less strictly also the tritium 
	experiments cut away the upper part of the allowed range 
	for the degenerate neutrino mass 
	($m_{com}= (0.14 - 3.6)\, eV$)
	could indicate 
	that the neutrino mass eigenvalues have the same CP parity 
\cite{KK-Cp-parity03}.

\begin{figure}[h]
\begin{center}
\includegraphics*[scale=0.4]{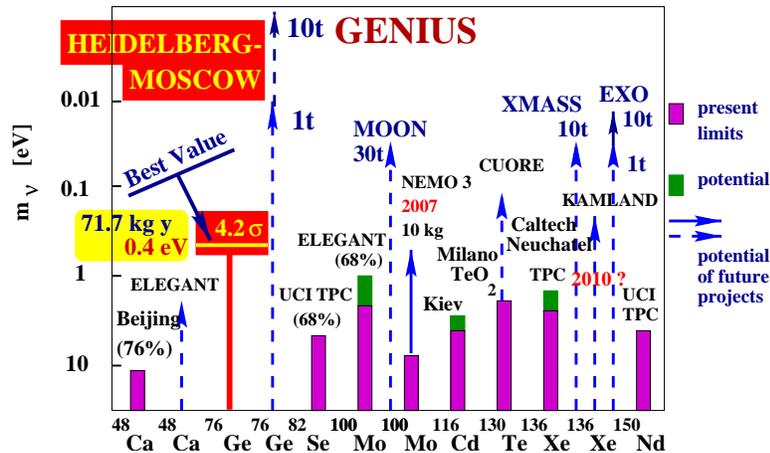}
\end{center}
\caption[]{
       Present sensitivity, and expectation for the future, 
       of the most promising $\beta\beta$ experiments. 
       Given are limits for $\langle m \rangle $, except 
	for the HEIDELBERG-MOSCOW experiment where 
	the measured {\it value} 
	is given (3$\sigma$ c.l. range and best value).
	Framed parts of the bars: present status; not framed parts: 
       future expectation for running experiments; solid and dashed lines: 
       experiments under construction or proposed, respectively. 
       For references see 
\protect\cite{KK60Y,KK02-PN,KK02-Found,KK-LowNu2,KK-NANPino00}.
\label{fig:Now4-gist-mass}}
\end{figure}

	Finally we briefly comment in section VIII about the possible future 
	of the field of double beta decay. 
	First results 
	from GENIUS-TF which has come into operation on May 5, 2003 
	in Gran Sasso with first in world 10\,kg of naked Germanium 
	detectors in liquid nitrogen 
\cite{TF-NIM03,GenTF-0012022,KK-Modul-NIM03}, 
	are discussed in another contribution to this report 
\cite{KK-IK-GTF-Valenc03}.



\section{Results Obtained in the Period August 2, 1990 
\protect\newline Until May 20, 2003.}

\begin{figure}[ht]
\hspace{.3cm}
\epsfysize=100mm\centerline{\epsffile{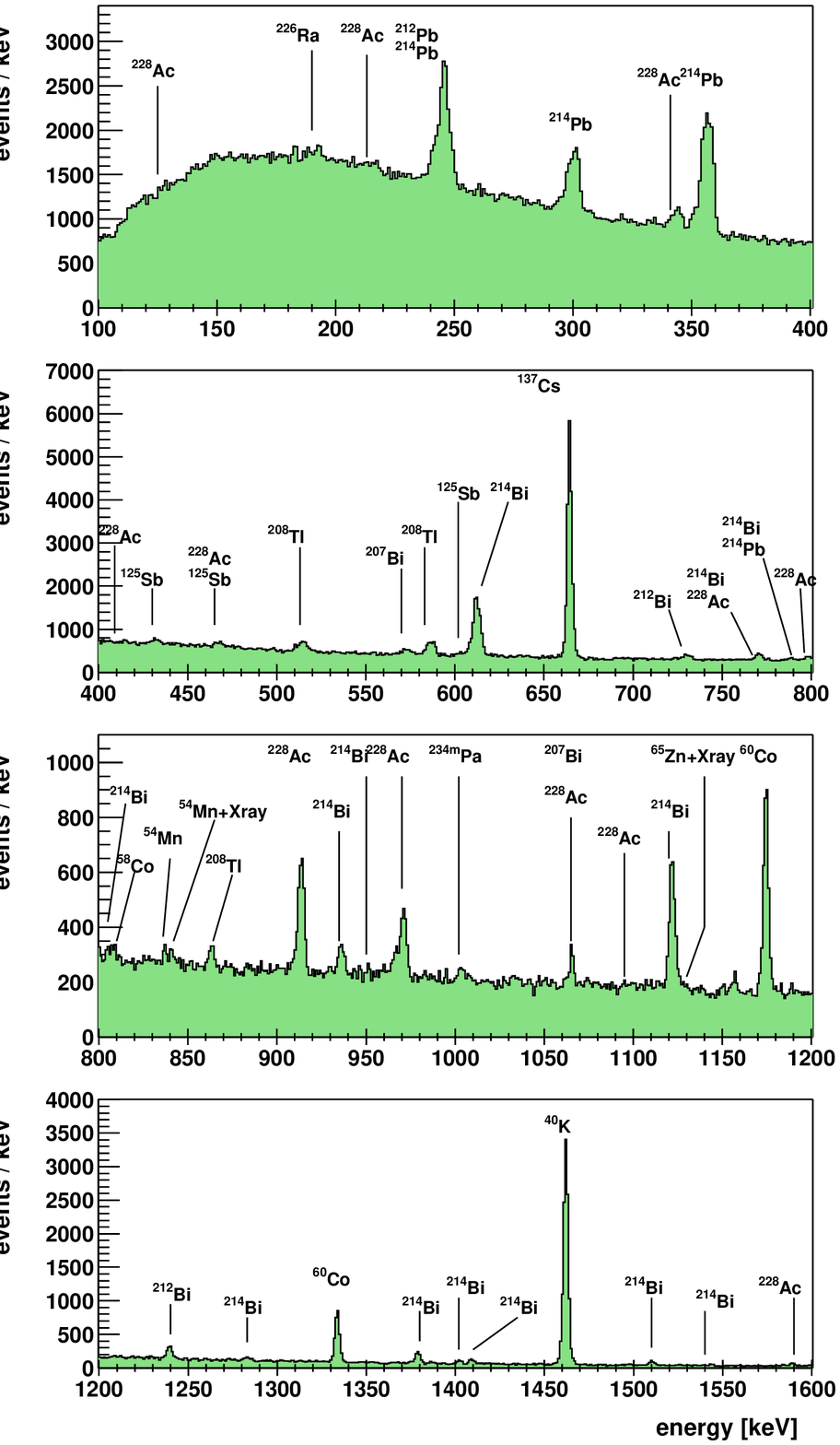}
\epsfysize=100mm\epsffile{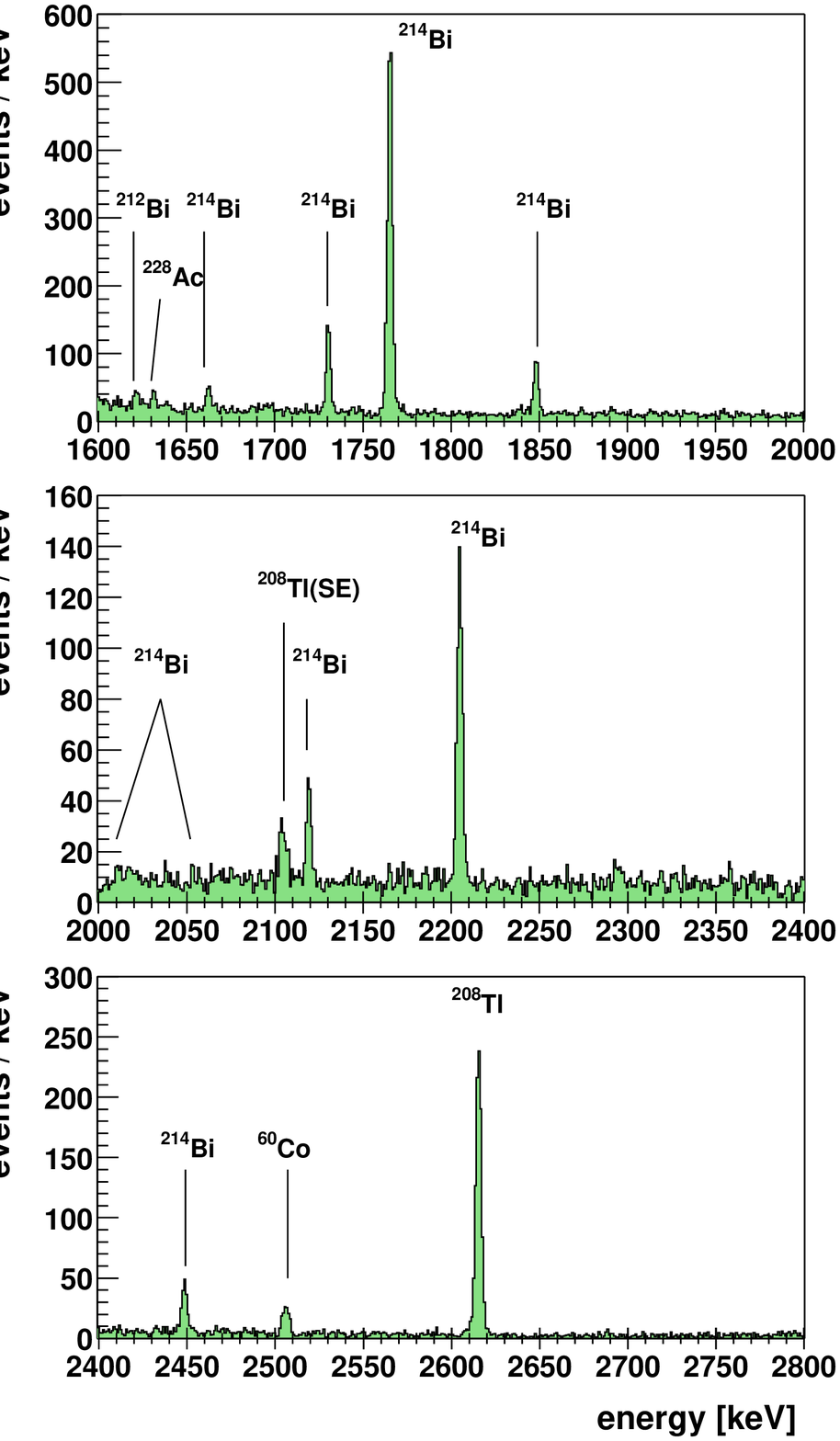}}
\caption[]{The total sum spectrum measured over the full energy range 
	(low-energy part (left), and higher energy part (right)) 
	of all five detectors (in total 10.96\,kg enriched 
	in $^{76}{Ge}$  to 86\%) - for the period 
	2 August 1990 to 20 May 2003.}
\label{fig:Low-HightAll90-03}
\end{figure}
\begin{figure}[ht]
\epsfysize=55mm\centerline{\epsffile{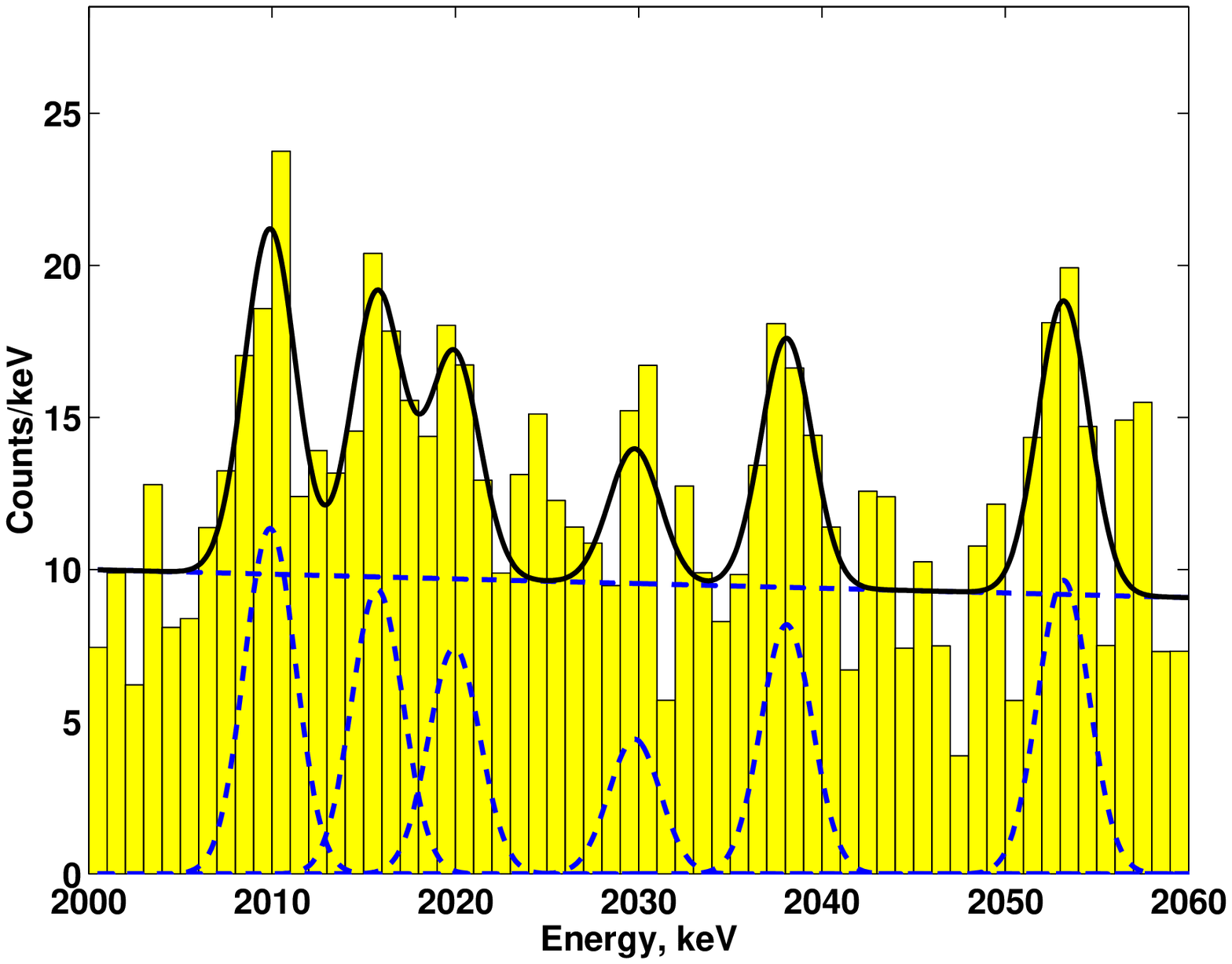}\hspace{0.7cm}\epsfysize=55mm\epsffile{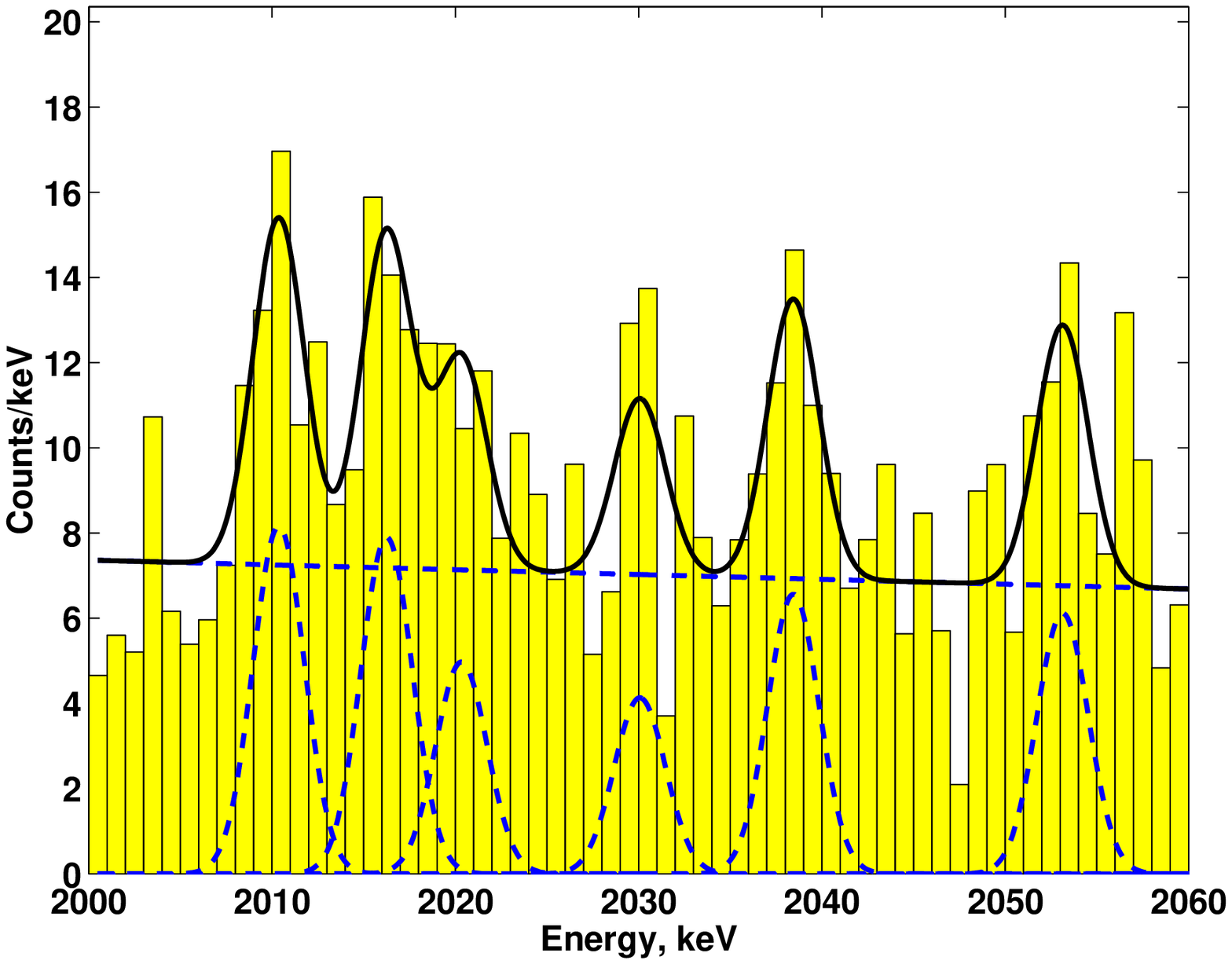}}
\caption[]{
    	The total sum spectrum of all five detectors 
	(in total 10.96\,kg enriched in $^{76}{Ge}$), for the period 
	August 1990 to May 2003 (71.7\,kg\,y) left, 
	and for the period November 1995-2003 (56.66\,kg\,y) 
	in the range 2000 - 2060\,keV 
	and its fit (see section 3.2).}
\label{fig:Sum90-03}
\end{figure}

	The status of present double beta experiments is shown in 
Fig.~\ref{fig:Now4-gist-mass}	
	and is extensively discussed in 
\cite{KK60Y}.	
	The HEIDELBERG-MOSCOW experiment using the largest source strength 
	of 11 kg of enriched $^{76}$Ge (enrichment 86$\%$) 
	in form of five HP Ge-detectors 
	is running since August 1990 
	in the Gran-Sasso underground laboratory 
\cite{KK60Y,KK02-Found,HDM01,KK02-PN,KK-StProc00,HDM97,Support-KK03-PL}. 
	We present here 
	in Figs.
\ref{fig:Low-HightAll90-03},\ref{fig:Sum90-03} 
	the results obtained with 
	three more years data, until May 20, 2003.
	Fig. 
\ref{fig:Low-HightAll90-03}
	shows the full spectrum, Fig. 
\ref{fig:Sum90-03}
	the range around the Q$_{\beta\beta}$ value.
	They correspond to a total measuring time of 71.7\,kg\,y.

	Fig.
\ref{fig:Sum90-03}
	shows that the line at Q$_{\beta\beta}$ is now - as 
	the Bi lines at 2010.7, 2016.7, 2021.8, 2052.9\,keV - 
	directly clearly seen, 
	while in our first results they had to be projected out 
	from the background by a peak search procedure 
\cite{KK02,KK02-PN,KK02-Found,Support-KK03-PL}.

	Earlier measurements 
	of Q$_{\beta\beta}$ by  
\cite{Old-Q-val,Q-val-Audi,Q-val-Ellis} 
	yielded  2040.71$\pm$0.52\,keV, 2038.56 $\pm$ 0.32\,keV and 
	2038.668$\pm$2.142\,keV. The precision measurement of 
\cite{New-Q-2001}
	yields 2039.006 (50)\,keV.

	The data have been analysed 
	with various statistical methods.
	We always process background-plus-signal data since 
	the difference between two Poissonian variables 
	does {\it not} produce a Poissonian distribution 
\cite{NIM99}. 
	This is important, but sometimes overlooked (see section 6).
	Analysis of the spectra 
	by nonlinear least squares method, 
	using

\clearpage

\noindent
	the Levenberg-Marquardt algorithm yields the fits, 
	shown in Fig. 
\ref{fig:Sum90-03}. 
	In these fits the peak positions, widths and intensities 
	of all lines 
	are determined simultaneously, and also the absolute level 
	of the background.
	The shape of the latter was chosen to be slightly 
	decreasing with energy, corresponding to the complete 
	simulation of the background performed in 
\cite{KK03}
	by GEANT4. E.g. in Fig.
\ref{fig:Sum90-03}, right,  
	the {\it fitted} background corresponds 
	to (55.94$\pm$3.92)\,kg\,y 
 	if extrapolated from the 
	background {\it simulated} in 
\cite{KK03}
	for the measurement with 49.59\,kg\,y  
	of statistics (see Fig. 
\ref{fig:bb_under}).
	This is almost {\it exactly} the statistical significance 
	of the present experiment (56.66\,kg y) 
	and thus a very nice proof of consistency. 
	Assuming a {\it constant} background in the range 2000 - 2060\,keV 
	or keeping also the {\it slope} of a linearly 
	varying background as a free parameter,
	yields very similar results. 
	Analysis with the Maximum Likelihood Method gives 
	results consistent with the above method.

	The signal at Q$_{\beta\beta}$ in the full spectrum 
	(the fit of Fig.
\ref{fig:Sum90-03}, right, 
	yields 
	2038.44$\pm$0.45\,keV), reaches a 4.2$\sigma$ 
	confidence level for the period 1990-2003, 
	and of 4.1$\sigma$ for the period 1995-2003 
	(for details we refer to 
\cite{NIM04-NEW-Res}). 
	A detailed description of the analysis 
	of the full data 1990-2003 will be given in the next Annual Report.

\vspace{-0.3cm}
\section{Measurements With a $^{214}{Bi}$ Source, 
\protect\newline Comparison With Other Ge-Experiments}

\begin{figure}[ht]
\epsfysize=40mm\centerline
{\epsffile{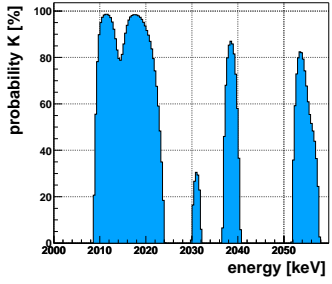}}
\caption[]{
	Scan for lines in the full spectrum 
	taken from 1995 - 2003  
	with detectors Nr. 1,2,3,4,5, 
	with the MLM method (see text). 
	The Bi lines at 2010.7, 2016.7, 2021.8 and 2052.9\,keV 
	are clearly seen, and in addition a signal at $\sim$ 2039\,keV. 
	}
\label{fig:Peak-Scan-Bay-12345-95-03Full}
\end{figure}

	By the peak 
	search procedure developped 
\cite{KK02-PN,KK02-Found}
	on basis of the Bayes and Maximum Likelihood Methods, 
	exploiting as
	important input parameters the experimental knowledge 
	on the shape and
	width of lines in the spectrum, weak lines of $^{214}$Bi 
	had been identified
	at the energies of 2010.7, 2016.7, 2021.6 and 2052.9\,keV 
	already in 
\cite{KK02,KK02-PN,KK02-Found,KK-StBr02}. 
	Though the lines with our improved statistics 
	and analysis are now clearly seen directly in the spectrum (Fig. 
\ref{fig:Sum90-03}), 
	we show for comparison the result of the peak search 
	procedure for the spectrum taken 1995-2003, in Fig. 
\ref{fig:Peak-Scan-Bay-12345-95-03Full}. 
	As usual, shown is the probability that there is a line 
	of correct width and of	Gaussian shape at a given energy, 
	assuming all the rest 
	of the spectrum as
	flat background (which is a highly conservative assumption).



	Concerning the intensities of these $^{214}{Bi}$ lines,  
	one has to note that the 2016\,keV line, as an E0 transition,
	can be seen only by coincident summing of the two successive lines
	$E=1407.98$\,keV and $E=609.316$\,keV. 
	Its observation proves that the $^{238}$U impurity from which
	it is originating, is located in the Cu cap of the detectors.

	We performed, in the first half of 2003, a {\it measurement} 
	of a $^{226}\rm{Ra}$ source with a high-purity 
	germanium detector 
\cite{Bi-KK03-NIM}. 
	The aim of this work 
	was to investigate the difference in the Bi spectra 
	when changing the 
	position of the source with respect to the detector, 
	and to verify the effect of TCS (true coincidence summing) 
	for the weak $^{214}{Bi}$ 
	lines seen in the \hdmo.

\begin{figure}[!htp]
\begin{center}
\includegraphics[width=8.cm,height=6.cm]{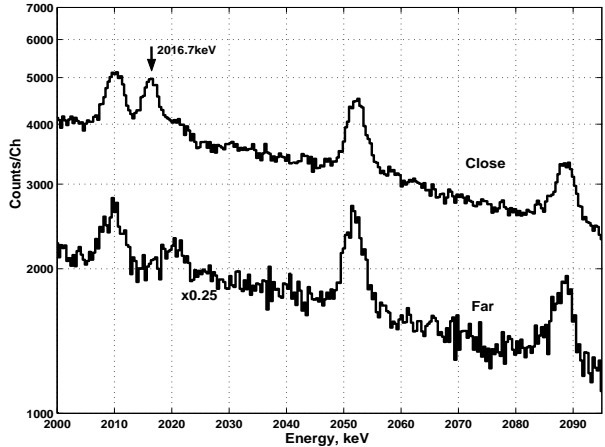}
\end{center}

\vspace{-0.5cm}
\caption[]{Measured $^{226}\rm{Ra}$ spectrum in the energy range
  	from 2000 to 2100\,keV. The upper spectrum corresponds to the close
  	geometry, the bottom spectrum to the far geometry.
  	The weak lines from $^{214}\rm{Bi}$ are
  	nicely visible, together with the effect of the true coincidence
  	summing at 2016.7\,keV (from 
\cite{Bi-KK03-NIM}).} 
\label{RaExper}
\end{figure}

	The activity of the $^{226}\rm{Ra}$ source was 
	95.2\,kBq. 
	The isotope $^{226}\rm{Ra}$ appears in the 
	$^{238}\rm{U}$ natural decay chain and from its decays also 
	$^{214}\rm{Bi}$ is produced. The $\gamma$-spectrum 
	of $^{214}\rm{Bi}$ 
	is clearly visible in the $^{226}\rm{Ra}$ measured spectrum 
	(see Fig.
\ref{RaExper}).
	We also performed a simulation of our measurement with the GEANT4 
	simulation tool and we find good agreement between the 
	simulation and the measurement 
\cite{Bi-KK03-NIM}. 
	The premature estimates of the Bi intensities given 
	in {\it Aalseth et al.},
	hep-ex/\-0202018 and Feruglio et al., 
	Nucl. Phys. {B 637} (2002), 345,
	{\it are incorrect}, because this long-known spectroscopic
	effect of true coincident summing 
\cite{gamma} 
	has not been taken into
	account, and also no simulation of the setup has been performed (for
	details see 
\cite{KK02-Found,KK-antw02,replay,Bi-KK03-NIM,Support-KK03-PL}).

	These Bi lines occur also in other investigations 
	of double beta decay.
	There are three other Ge experiments 
	which have looked 
	for double beta decay of $^{76}$Ge. 
	First there is the experiment by Caldwell et al. 
\cite{Caldw91}, 
	using natural Germanium detectors (7.8\% abundance 
	of $^{76}$Ge, compared to 86\% in
	the HEIDELBERG-MOSCOW experiment). 
	This was the most sensitive {\it natural} Ge
	experiment. With their background a factor of 9 higher than in the
	HEIDELBERG-MOSCOW experiment and their measuring time 
	of 22.6 kg\,years,
	they had a statistics of the background by a factor of almost four
	\mbox{l a r g e r} than in the HEI\-DELBERG-MOS\-COW experiment. 
	This gives useful
	information on the composition of the background.

\begin{figure}[h]

\vspace{-0.4cm}
\begin{center}
\includegraphics*[scale=0.3]{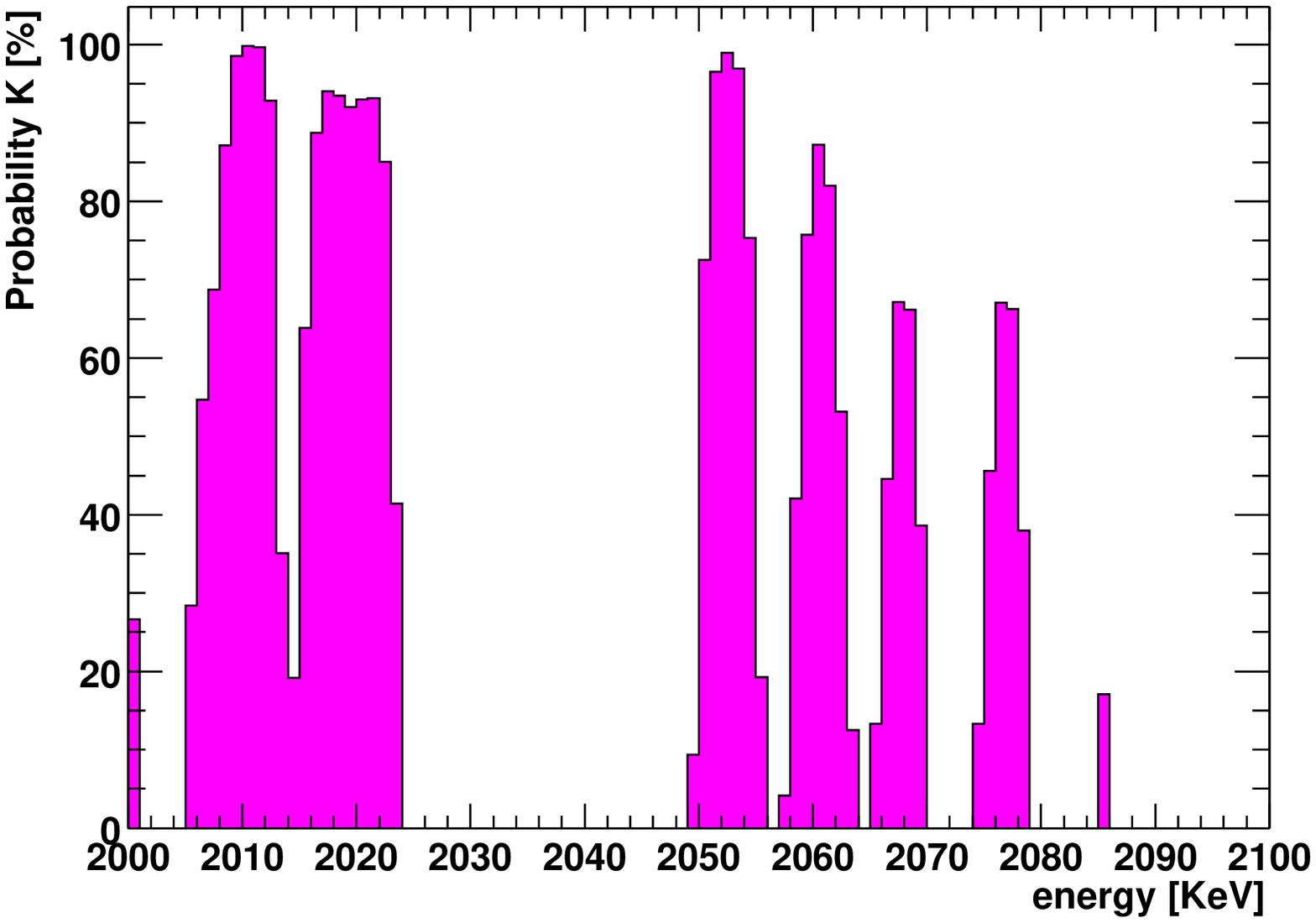}
\includegraphics*[scale=0.3]{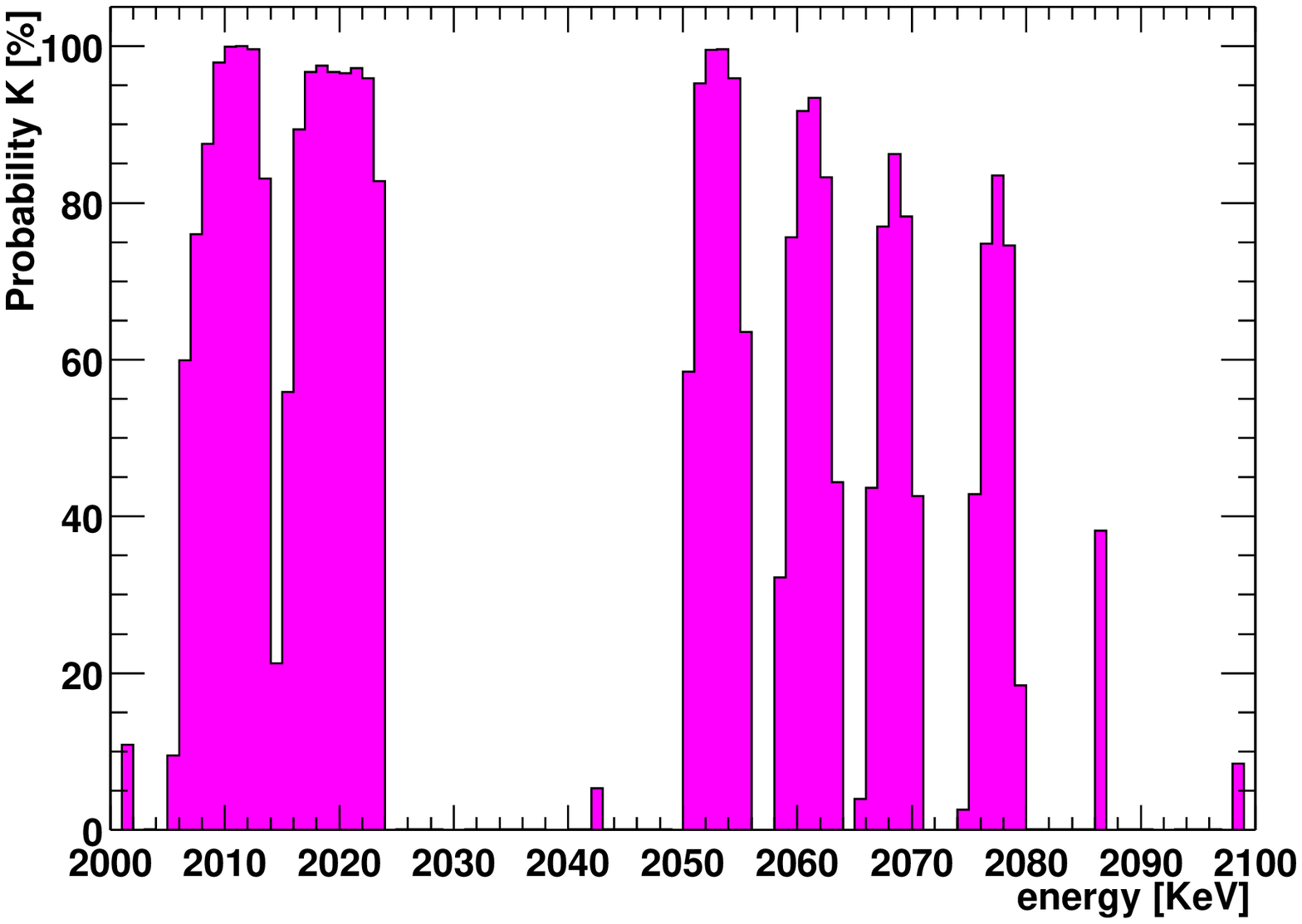}
\end{center}

\vspace{-0.5cm}
\caption{
	Result of the peak-search procedure
	performed for the UCBS/LBL spectrum 
\protect\cite{Caldw91} 
	(left: Maximum Likelihood method, right: Bayes method). 
	On the y axis the probability 
	of having a line at the corresponding energy 
	in the spectrum is shown (from 
\protect\cite{Support-KK03-PL,Backgr-KK03-NIM}).
\label{fig:figCaldwell}}
\end{figure}

\begin{figure}[th]

\vspace{-0.5cm}
\begin{center}
\includegraphics*[scale=0.27]{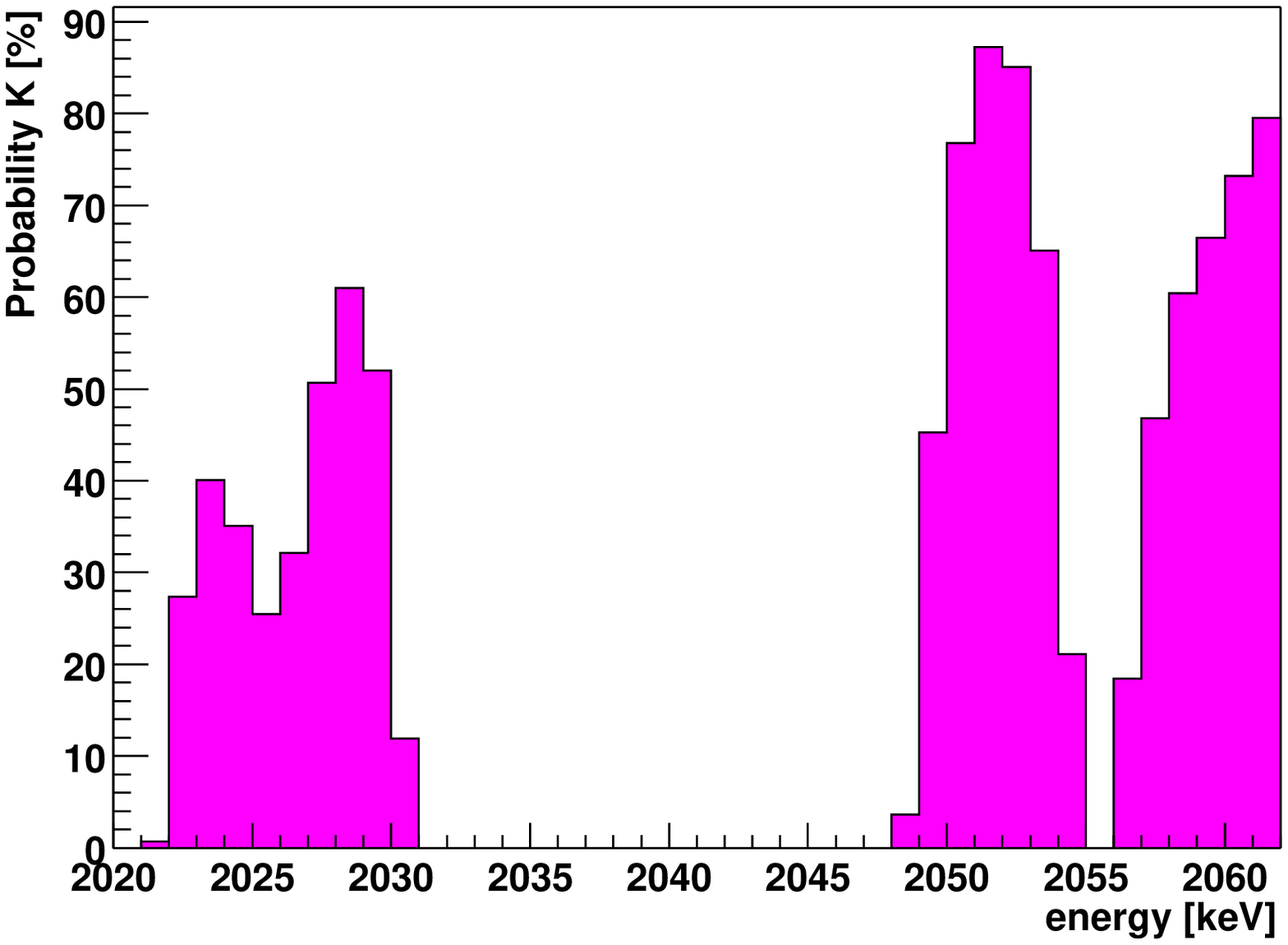}
\includegraphics*[scale=0.27]{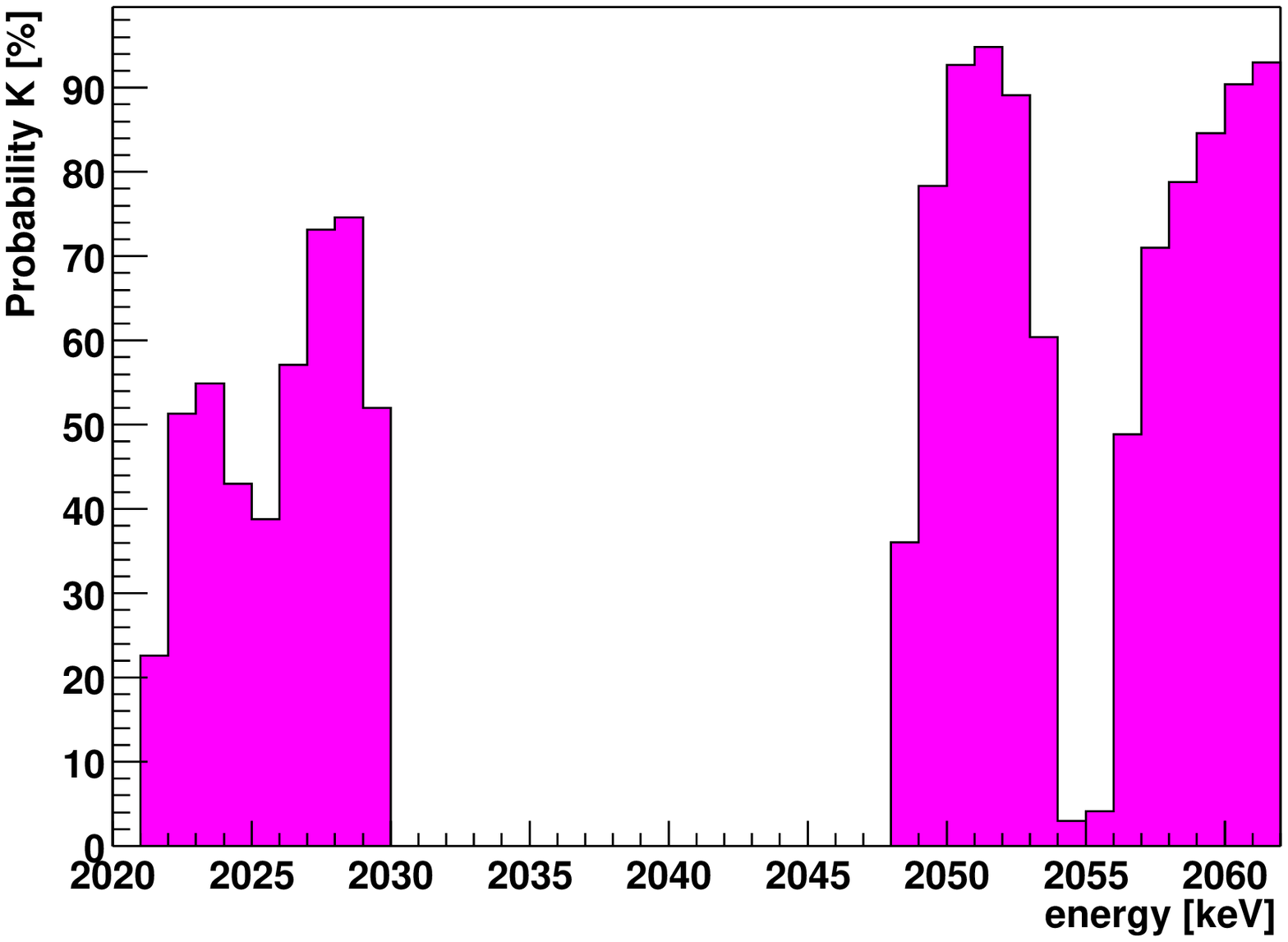}
\end{center}

\vspace{-0.5cm}
\caption{Result of the peak-search procedure
	performed for the ITEP/YePI spectrum 
\protect\cite{Kirpichn} 
	(from 
\protect\cite{Support-KK03-PL,Backgr-KK03-NIM}).
\label{fig:figITEP}}

\vspace{-0.3cm}
\begin{center}
\includegraphics*[scale=0.25]{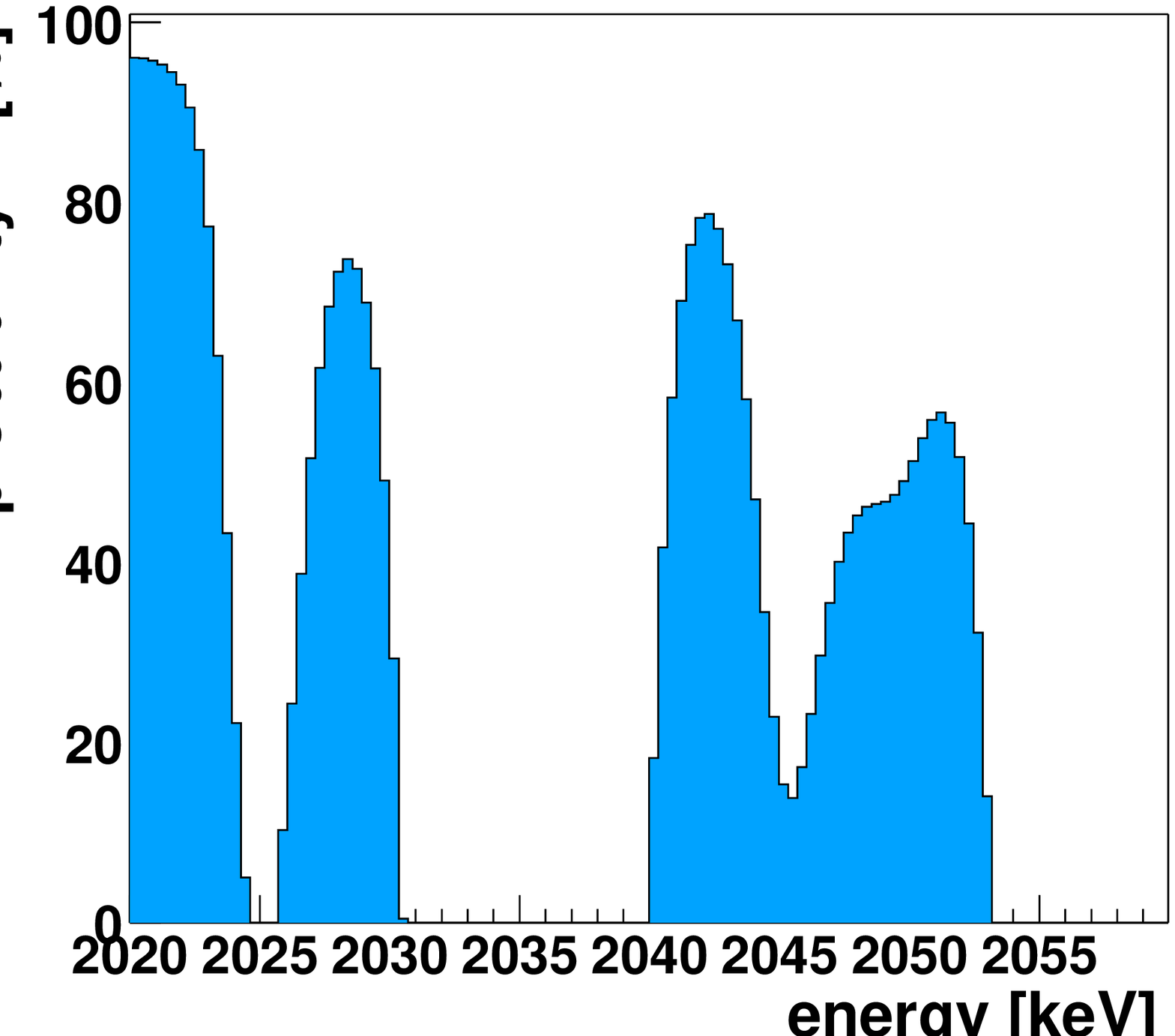}
\includegraphics*[scale=0.25]{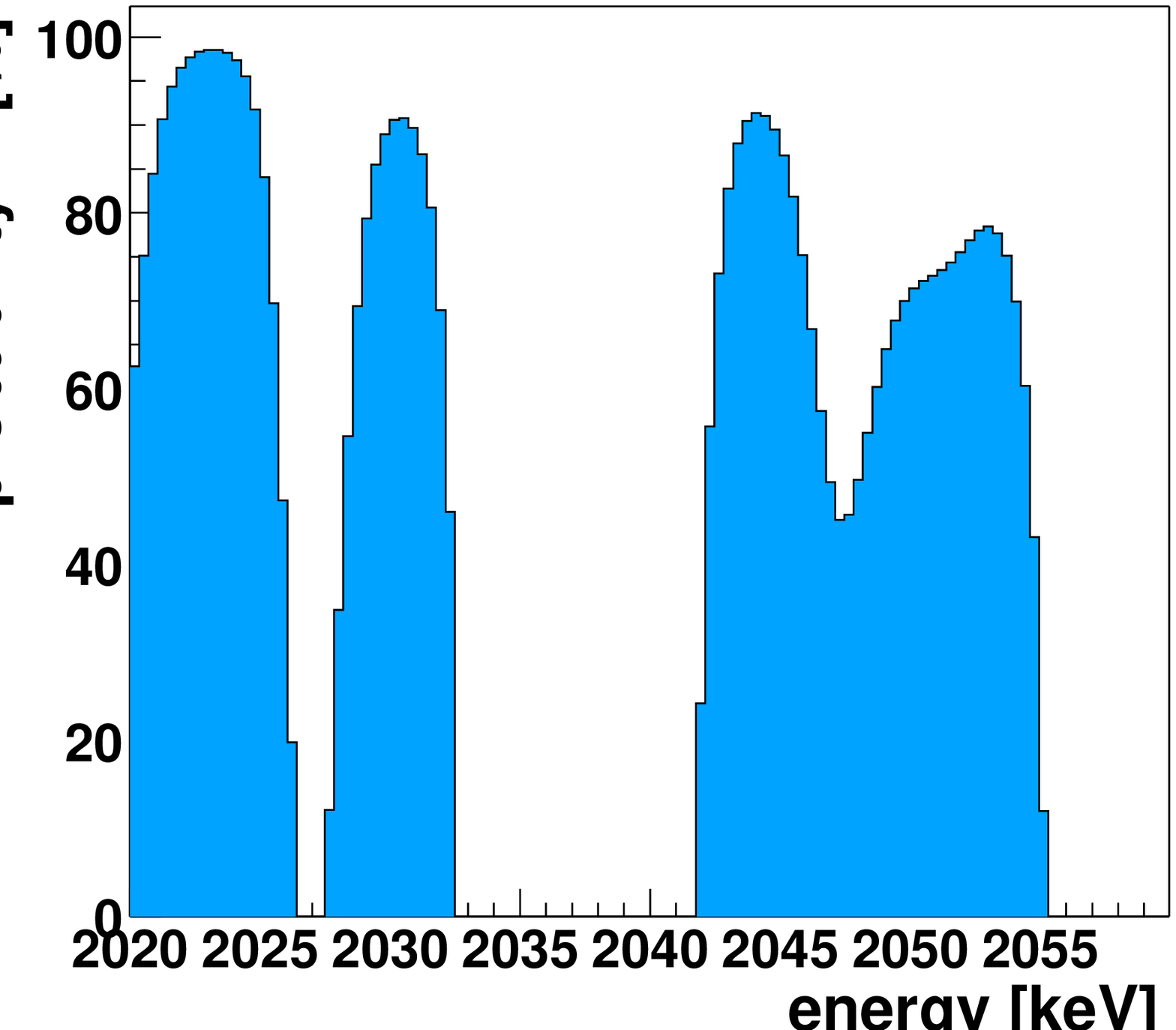}
\end{center}

\vspace{-0.5cm}
\caption{Result of the peak-search procedure
	performed for the IGEX spectrum 
\protect\cite{DUM-RES-AVIGN-2000}.
	Left: Maximum Likelihood method, right: Bayes method. 
	On the y axis the probability 
	of having a line at the corresponding energy 
	in the specrtum is shown (from 
\protect\cite{Support-KK03-PL,Backgr-KK03-NIM}).
\label{fig:figIGEX}}


\begin{center}
\includegraphics*[scale=0.4]{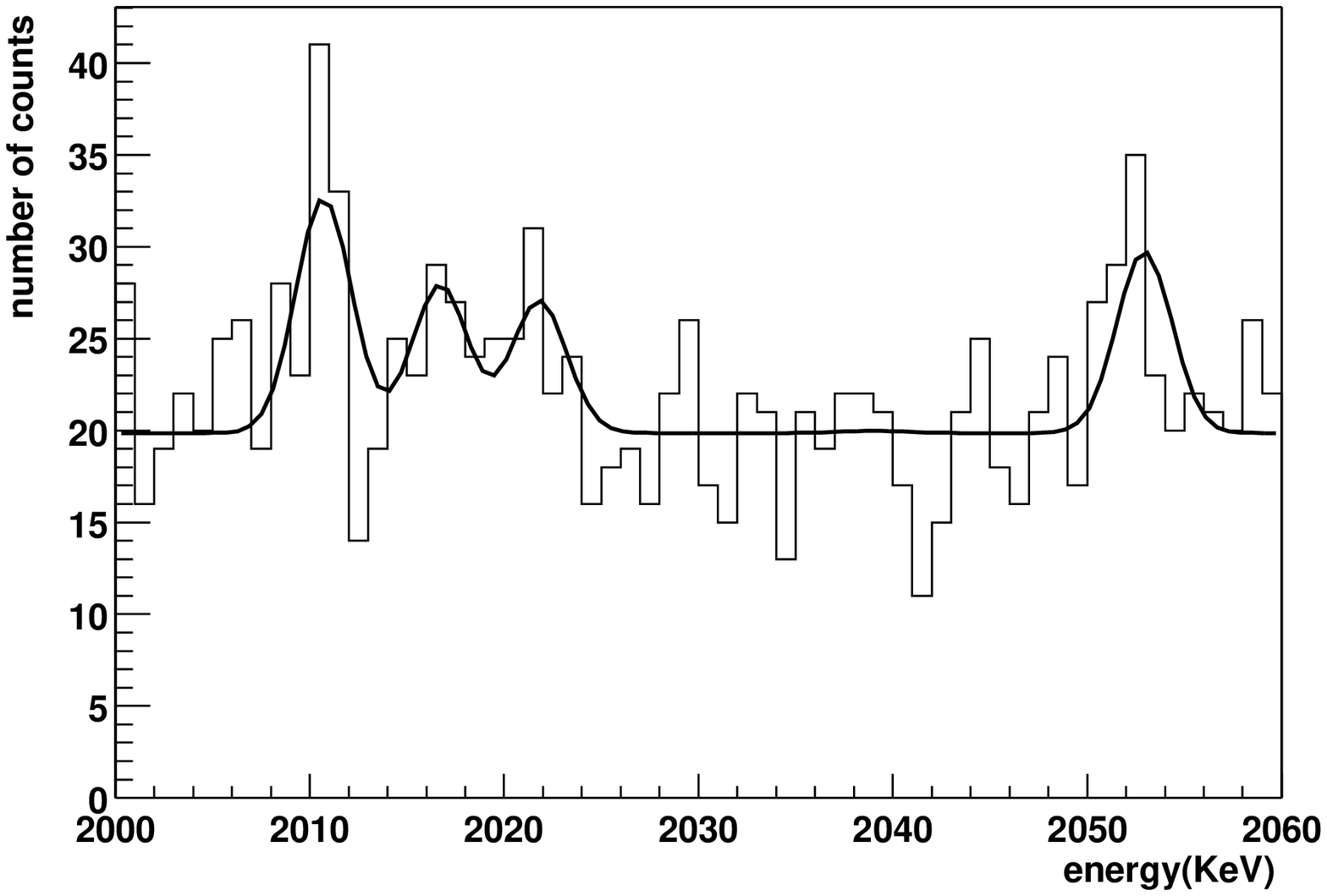}
\end{center}

\vspace{-0.5cm}
\caption{ 
	Analysis of the spectrum measured by D. Caldwell et al. 
\protect\cite{Caldw91},
       with the Maximum Likelihood Method, in the energy 
        range 2000-2060\,keV 
        assuming lines at 2010.7, 2016.7, 2021.6, 2052.9,
	2039.0\,keV.        
        No indication for a signal at 2039\,keV is observed in this case 
	(see 
\protect\cite{Support-KK03-PL}).
\label{fig:specCaldwell}}
\end{figure}

	Applying the same method of peak search as used in 
Fig. 
\ref{fig:Peak-Scan-Bay-12345-95-03Full}, 
	yields (see also 
\cite{Support-KK03-PL,Backgr-KK03-NIM}) 
	indications for peaks essentially at the same energies as in
Fig. \ref{fig:Peak-Scan-Bay-12345-95-03Full} 
(see Fig. \ref{fig:figCaldwell}). 		
	This shows that these peaks are not fluctuations. In particular
	it sees the 2010.78, 2016.7, 2021.6 and 
	2052.94\,keV $^{214}$Bi 
	lines, but~~ a l s o~~  the unat\-tri\-buted lines 
	at higher energies.  
	It finds, however, n o  line at 2039\,keV. 
	This is consistent with the
	expectation from the rate found 
	in the HEIDELBERG-MOSCOW experiment.
	About 29 identified events observed during 1990-2003 
	in the latter correspond to 0.7  expected
	events in the Caldwell experiment, because of the use 
	of non-enriched material and the shorter measuring time. 
	Fit of the Caldwell spectrum
	allowing for the $^{214}$Bi lines and a 2039\,keV line 
	yields 0.4\,events for the latter (see 
\cite{KK02-Found} and Fig. 
\ref{fig:specCaldwell}).


	The first experiment using enriched (but not high-purity) 
	Germanium 76
	detectors was that of Kirpichnikov and coworkers 
\cite{Kirpichn
}. 
	These authors show only the energy range between 2020 
	and 2064\,keV of their measured spectrum.
	The peak search procedure finds also here indications of lines
	around 2028\,keV and 2052\,keV (see Fig. 
\ref{fig:figITEP}), 
	but \mbox{n o t} any indication of a line at 2039\,keV. 
	This is consistent with the expectation, because for their low
	statistics of 2.95 kg\,y they would expect here (according to
	HEIDELBERG-MOSCOW) 1.1\,counts.

\clearpage

\clearpage

	Another experiment (IGEX) used between 6 and 8.8\,kg 
	of enriched $^{76}$Ge,
	but collected since beginning of the experiment in the early nineties
	till shutdown in 1999 only 8.8 kg\,years of statistics 
\cite{DUM-RES-AVIGN-2000}.
	The authors of 
\cite{DUM-RES-AVIGN-2000} 
	unfortunately show only the range 2020 to 2060\,keV 
	of their measured spectrum in detail. Fig. 
\ref{fig:figIGEX}
	shows the result of our peak scanning 
	of this range. 
	Clear indications are seen for
	the lines at 2021 and 2052\,keV, but also 
	of the unidentified structure around 2030\,keV. 
	Because of the conservative assumption on the background
	treatment in the scanning procedure (see above) there 
	is no chance to see
	a signal at 2039\,keV because of the 'hole' in the background of that
	spectrum (see Fig. 1 in 
\cite{DUM-RES-AVIGN-2000}). 
	With some good will one might see, however,
	an indication of $\sim$3\,events here, consistent 
	with the expectation of the
	HEIDELBERG-MOSCOW experiment 
	\protect\newline of $\sim$ 2.6\,counts.

\vspace{-0.3cm}
\section{Statistical Features: \protect\newline Sensitivity of Peak Search, 
	Analysis Window}

\begin{figure}[h]

\vspace{-0.3cm}
\begin{center}
\includegraphics*[scale=0.3]{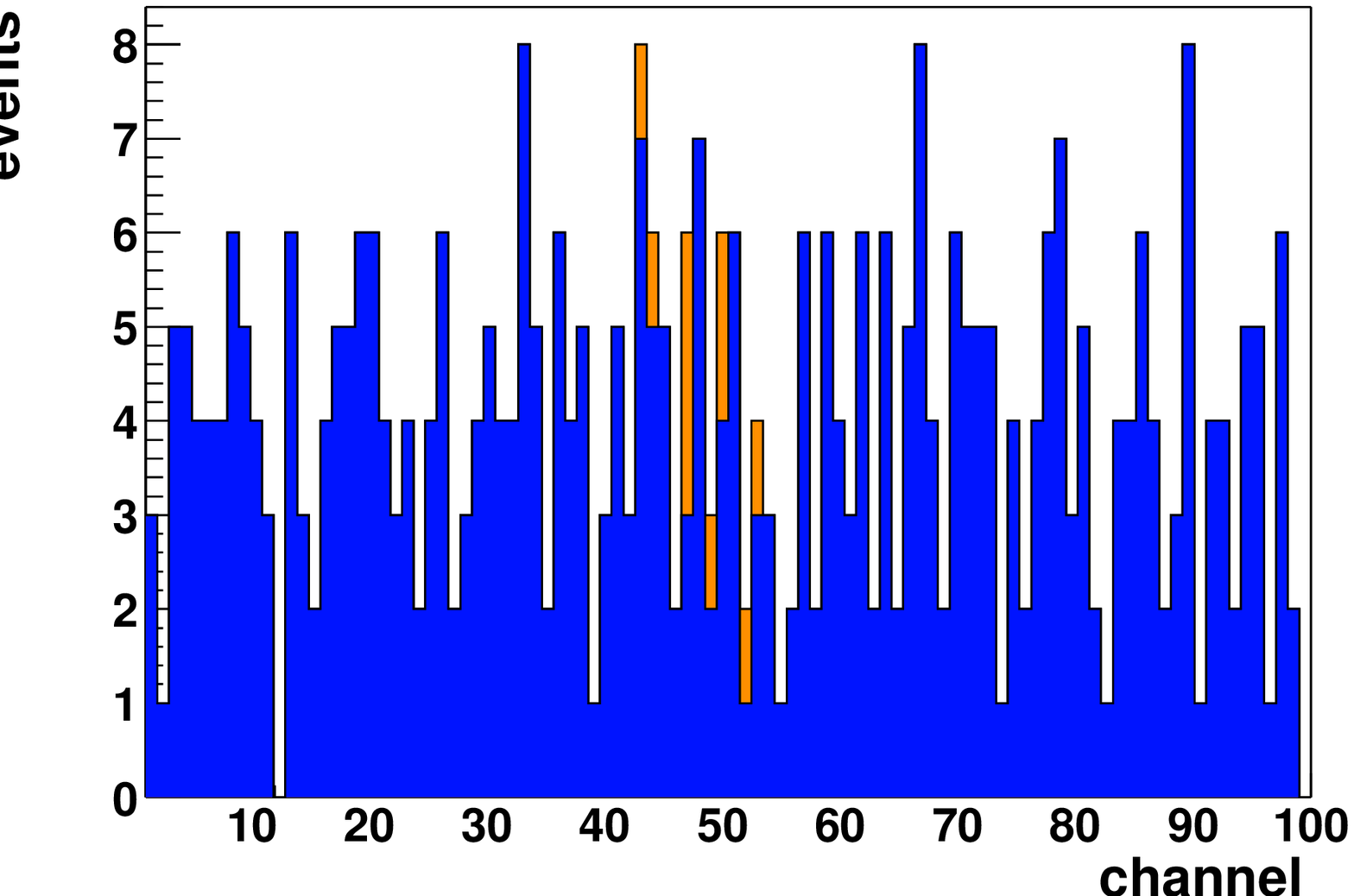}
\includegraphics*[scale=0.3]{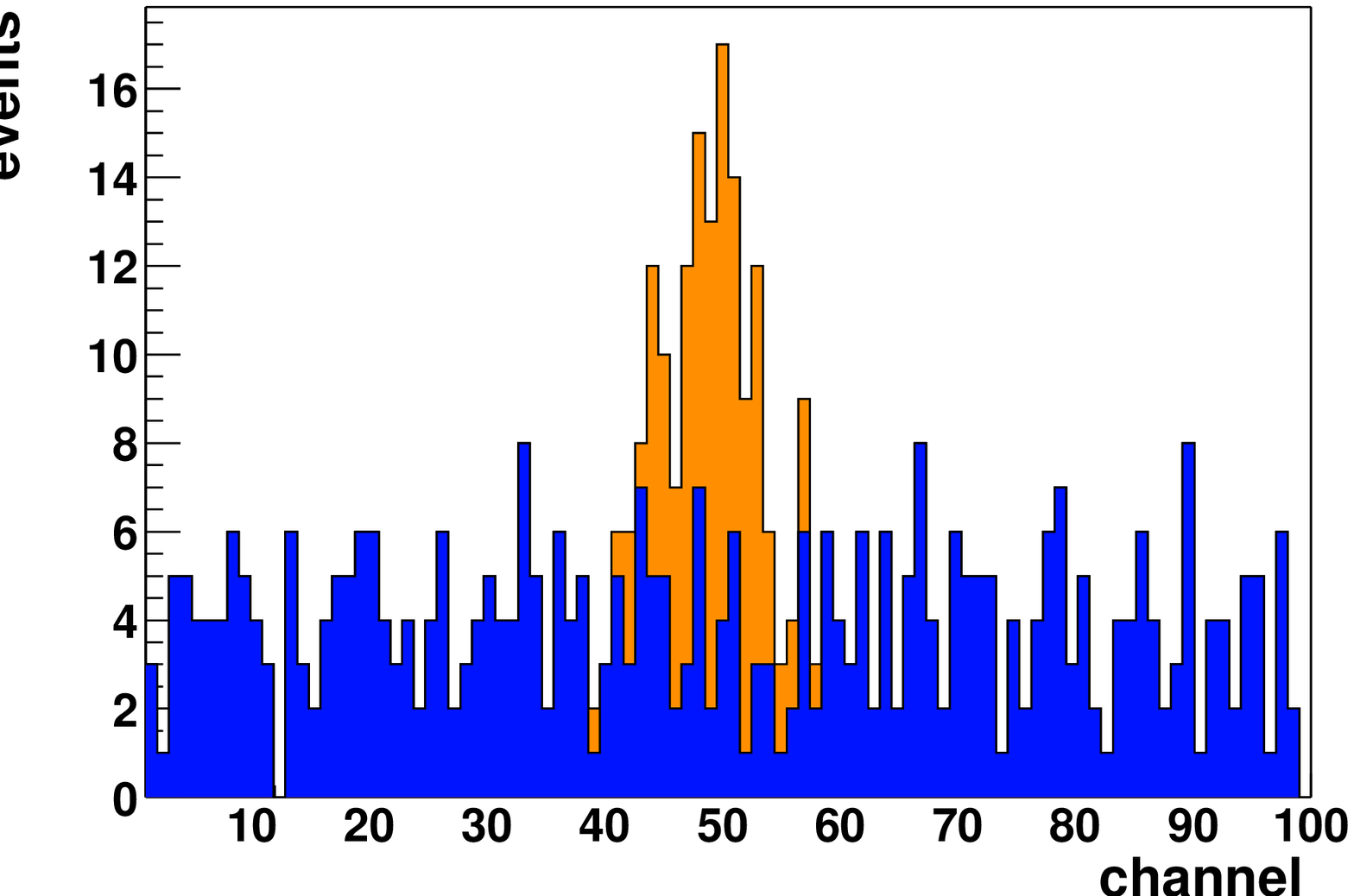}
\end{center}

\vspace{-0.5cm}
\caption{\rm \small 
	Example of a random-generated spectrum with a Poisson distributed
	background with 4.0\,events per channel and a Gaussian 
	line centered in channel
	50 (line-width corresponds to a standard-deviation of $\sigma=4.0$
	channels).
	The left picture shows a spectrum with a line-intensity of 10\,events,
	the right spectrum a spectrum with a line-intensity of 100\,events.
	The background is shown dark, the events of the line bright (see  
\protect\cite{Backgr-KK03-NIM}).
\label{fig:picSpecKu}}
\end{figure}

	For historical reasons, 
	at this point it may be useful to demonstrate the potential 
	of the peak search procedure used in 
\cite{KK02,KK02-PN,KK02-Found}. 
	Fig. 
\ref{fig:picSpecKu} 
	shows a spectrum with Poisson-generated background of
	4 events per channel and a Gaussian line with width (standard
	deviation) of 4 channels centered at channel 50, with intensity of 10
	(left) and 100 (right) events, respectively.
	Fig. 
\ref{fig:picPrior},
	shows the result of the analysis of spectra of
	different line intensity with the Bayes method (here Bayes 
	1-4 correspond to different choice of the prior distribution:
	(1) $\mu(\eta)=1$ (flat), (2) $\mu(\eta) = 1/\eta$, 
	(3) $\mu(\eta) = 1/\sqrt{\eta}$,
	(4) Jeffrey's prior) and the Maximum Likelihood Method.
	For each prior 1000 spectra have been generated with equal 
	background and equal line intensity using random number 
	generators available at CERN 
\cite{Random}.
	The average values of the best values agree (see Fig. 
\ref{fig:picPrior}) 
	very well with the known intensities also for very low 
	count rates (as in Fig.
\ref{fig:picSpecKu}, 
	left).



\begin{figure}[h]
\begin{center}
\includegraphics*[scale=0.35]{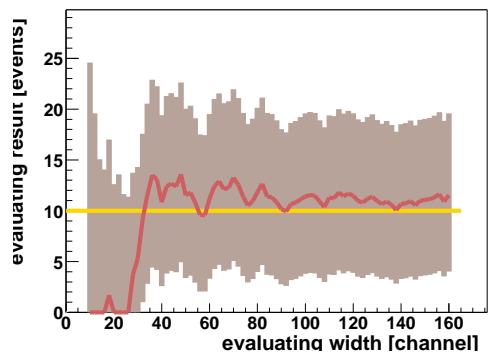}
\end{center}

\vspace{-0.5cm}
\caption{Result of an analysis as function of the evaluation width.
	The used spectrum consists of a Poisson distributed background with 4
	events per channel, and a line of 10 events (see Fig. 
\ref{fig:picSpecKu},
	left part). 
	The dark area corresponds to a 68.3\% confidence area with the dark
	line being the best value.
	Below an evaluation width of 35 channels the result becomes unreliable,
	above 35 channels the result is stable (see also 
\cite{Support-KK03-PL,Backgr-KK03-NIM}).
\label{picSingleDev}}
\end{figure}

	In Fig. 
\ref{fig:picWH1} 
	we show two simulations of a Gaussian line of 15\,events, 
	centered at channel 50, again with width (standard deviation) 
	of 4 channels, on a Poisson-distributed background 
	with 0.5\,events/channel. The figure gives an
	indication of the possible degree of deviation of the energy of the
	peak maximum from the transition energy,  on the level of statistics
	collected in experiments like the HEIDELBERG-MOSCOW experiment 
	(here one channel corresponds to 0.36\,keV). 
	This should be kept in mind.


\begin{figure}[h]
\begin{center}
\includegraphics*[scale=0.3]{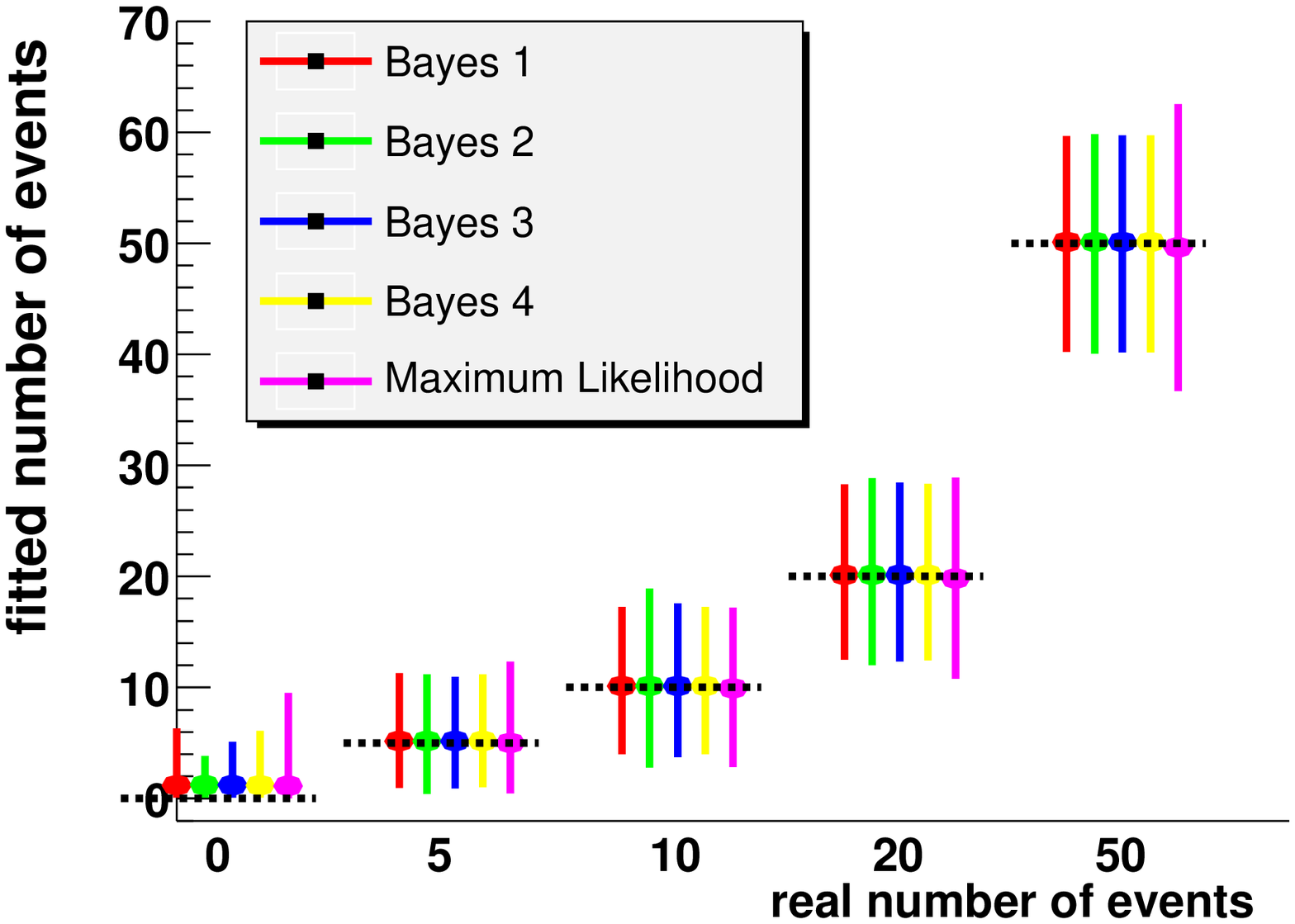}
\includegraphics*[scale=0.3]{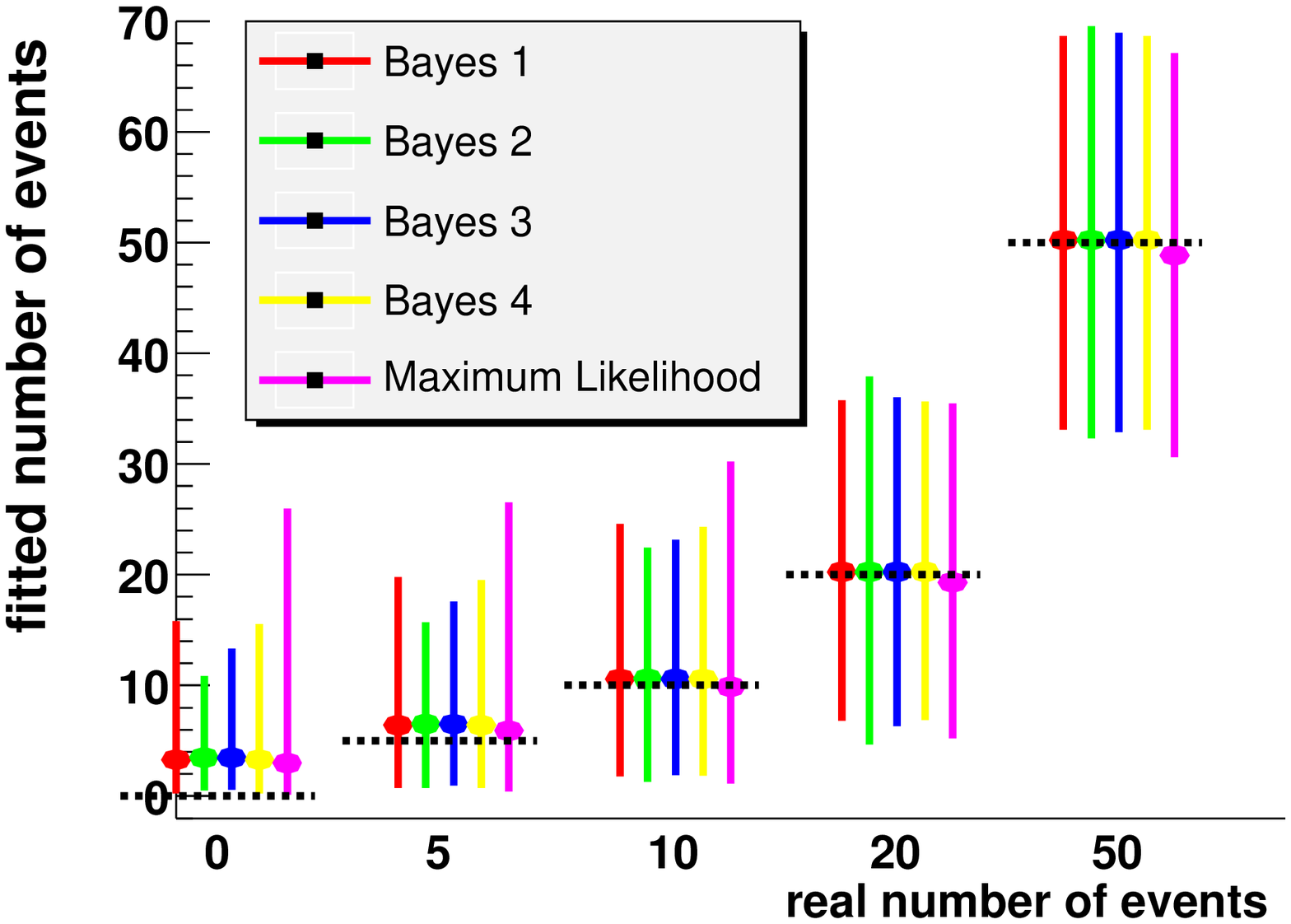}
\end{center}

\vspace{-0.5cm}
\caption{\rm \small 
	Results of analysis of random-number generated spectra, 
	using Bayes and Maximum Likelihood method (the first 
	one with different prior distributions).
	For each number of events in the simulated line, 
	shown on the x-axis, 1000 random generated 
	spectra were evaluated with the five given methods.
	The analysis on the left side was performed with an Poisson
	distributed background of 0.5\,events per channel, the background for
	the spectra on the right side was 4.0\,events per channel.
	Each vertical line shows the mean value of the calculated best values
	(thick points) with the 1$\sigma$ error area.
	The mean values are in good agreement with the expected values 
	(horizontal black dashed lines) (see  
\protect\cite{Support-KK03-PL,Backgr-KK03-NIM}).
\label{fig:picPrior}}
\end{figure}



\begin{figure}[h]
\begin{center}
\includegraphics*[scale=0.3]{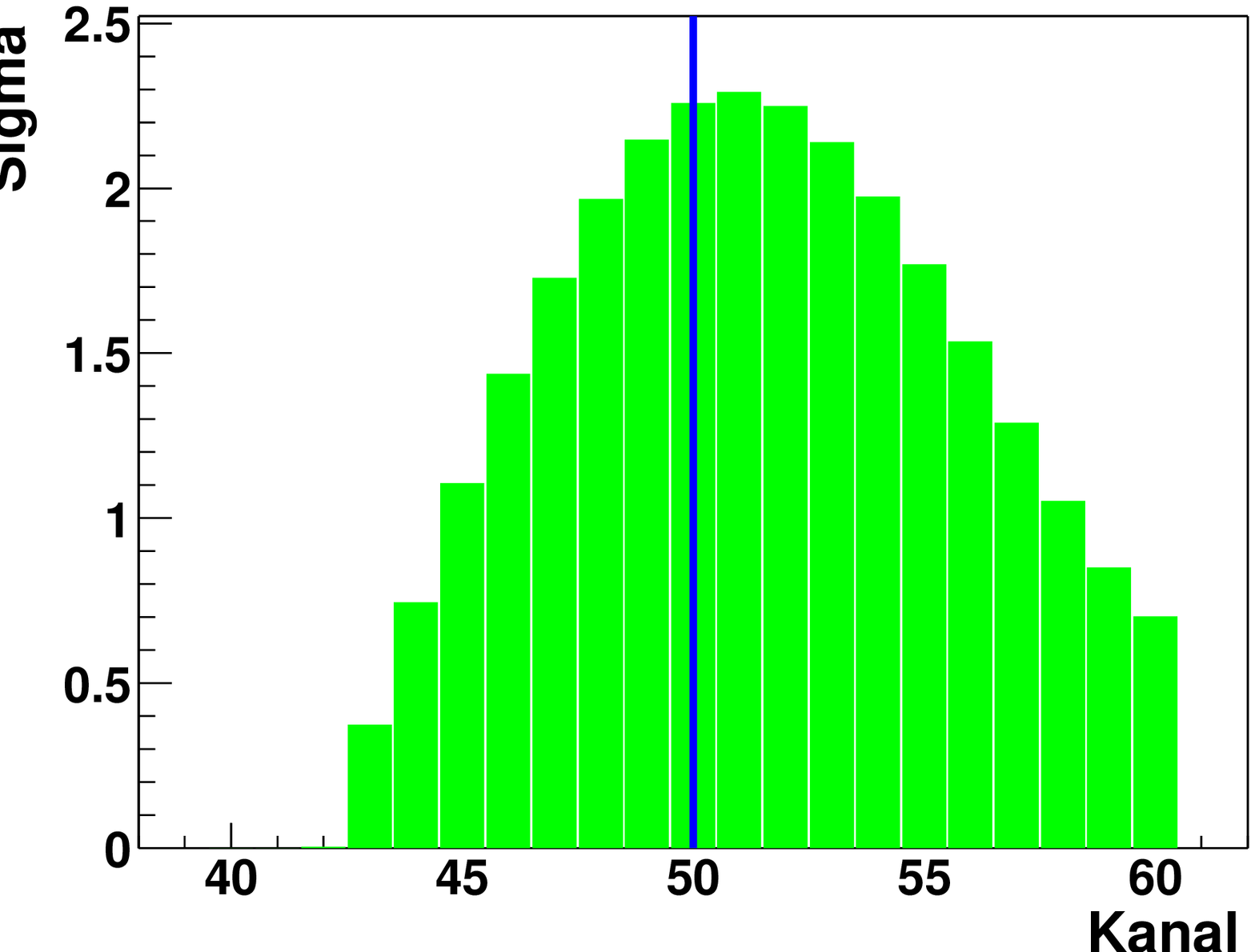}
\includegraphics*[scale=0.3]{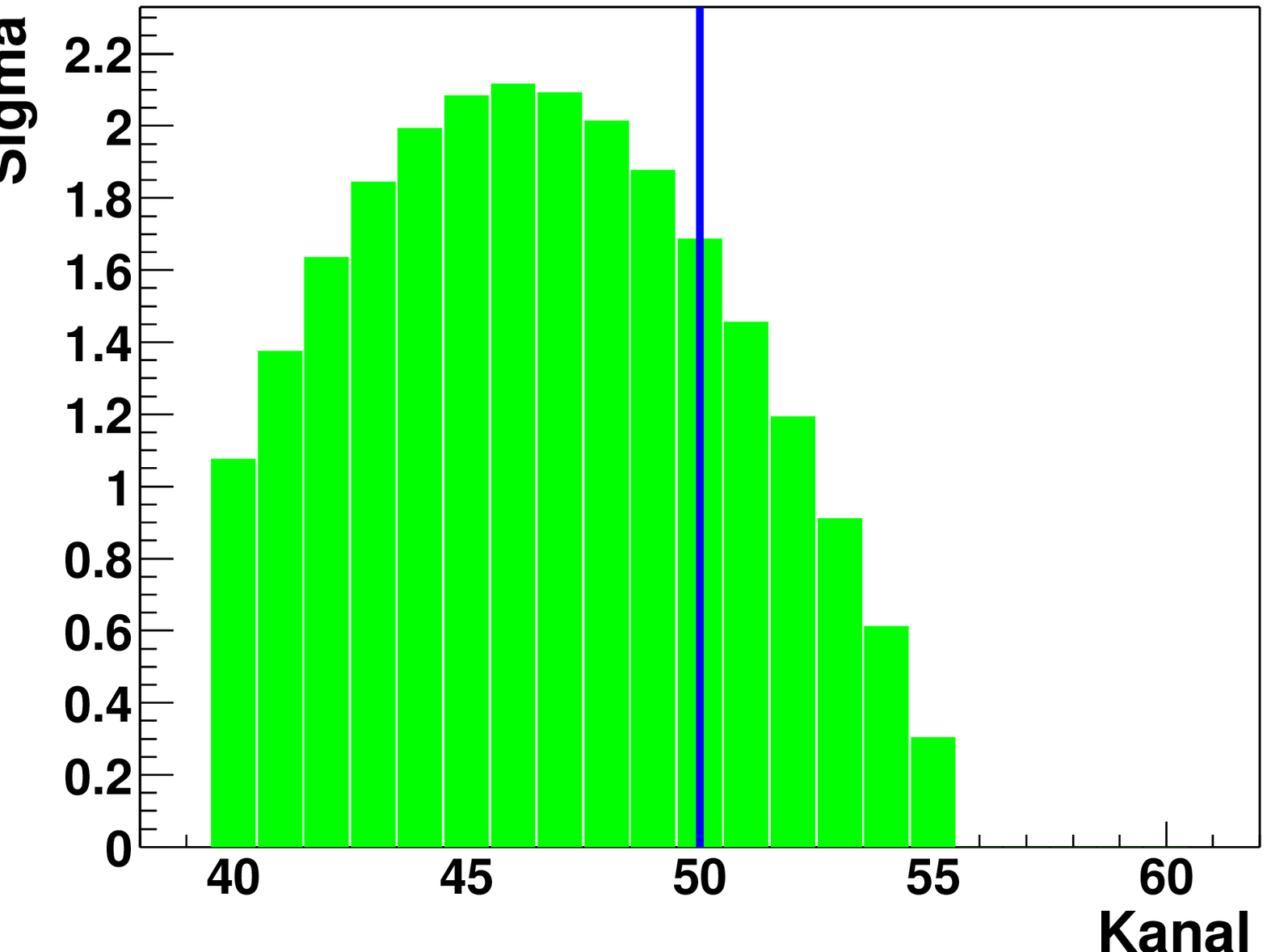}
\end{center}

\vspace{-0.5cm}
\caption{ \rm \small 
	Two spectra with a Poisson-distributed background and a
	Gaussian line with 15 events centered in channel 50 (with a width
	(standard-deviation) of 4.0 channels) created with different 
	random numbers. 
	Shown is the result of the peak-scanning of the spectra.
	In the left picture the maximum of the probability corresponds well
	with the expected value (black line) whereas in the right
	picture a larger deviation is found. 
	When a channel corresponds  to 0.36\,keV the deviation in the right
	picture is $\sim$ 1.44\,keV (see  
\protect\cite{Support-KK03-PL,Backgr-KK03-NIM}).
\label{fig:picWH1}}
\end{figure}


	The influence of the choice of the energy range 
	of the analysis around $Q_{\beta\beta}$
	has been thoroughly discussed in 
\cite{KK02-PN,KK02-Found}. 
	Since erroneous ideas about this
	point are still around, 
	let us remind of the analysis given in 
\cite{KK02-PN,KK02-Found,Backgr-KK03-NIM,Support-KK03-PL}
	which showed that a reliable result is
	obtained for a range of analysis of not smaller than 35 channels
	(i.e. $\pm$18 channels) - one channel corresponding 
	to 0.36\,keV in the
	HEIDELBERG-MOSCOW experiment (see Fig. 
\ref{picSingleDev}).
	This is an important result, since it is, 
	in case of a weak signal, {\it of course} important 
	to keep the range of analysis as  \mbox{s m a l l} 
	as possible, to avoid to include
	lines in the vicinity of the weak signal into the background 
	(see, e.g. Fig. 9 in 
\cite{XMASS03}).
	This unavoidably occurs when e.g. proceeding as suggested 
	in F. Feruglio et al., hep-ph/0201291 and 
	Nucl. Phys. B 637 (2002) 345-377, 
	Aalseth et. al., hep-ex/0202018 and 
	Mod. Phys. Lett. A 17 (2002) 1475,
	Yu.G. Zdesenko et. al., Phys. Lett. B 546 (2002) 206, 
	A. Ianni, in Press NIM 2004.
	The arguments given in those papers are therefore {\it incorrect}. 
	Also Kirpichnikov, who states 
\cite{Kirpichn
} 
	that his analysis finds a 2039\,keV signal in
	the HEIDELBERG-MOSCOW spectrum on a 4 sigma confidence 
	level (as we also see it) makes this mistake, 
	when analysing the pulse shape spectrum.

	The above discussion is now in this context only 
	of historical interest, 
	since with the better statistics we have now, we can  
	analyze simultaneously a large energy range (as shown in Fig. 
\ref{fig:Sum90-03}).


\vspace{-0.3cm}
\section{Simulation with GEANT4}

\begin{figure}[h]
\begin{center}
\includegraphics*[scale=0.55]{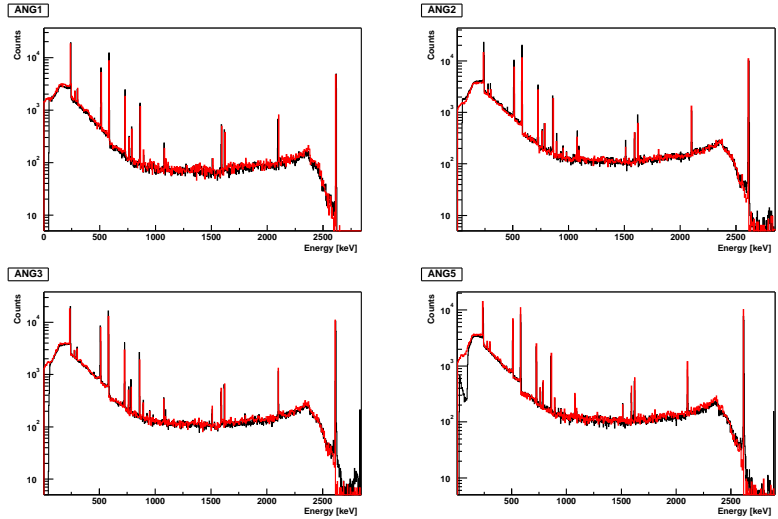}
\end{center}

\vspace{-0.5cm}
\caption{Comparison of the
	measured data (black line, November 1995 to April 2002) 
	and simulated
	spectrum (red line) for the detectors Nrs. 1,2,3 and 5 for a
	$^{232}$Th source spectrum.
	The agreement of simulation and measurement is excellent (from 
\protect\cite{KK03
}).
\label{fig:picUnderTotal}}
\end{figure}

	Finally the background around $Q_{\beta\beta}$ 
	will be discussed from the side of
	simulation. A very careful new simulation of 
	the different components of
	radioactive background in the HEIDELBERG-MOSCOW experiment has been
	performed by a new Monte Carlo program based on GEANT4 
\cite{KK03}.
	This simulation uses a new event generator for simulation 
	of radioactive
	decays basing on ENSDF-data and describes the decay of arbitrary
	radioactive isotopes including alpha, beta and gamma emission 
	as well as
	conversion electrons and X-ray emission. Also included 
	in the simulation is
	the influence of neutrons in the energy range from thermal to high
	energies up to 100\,MeV on the measured spectrum. 
	Elastic and inelastic
	reactions, and capture have been taken into account, 
	and the corresponding
	production of radioactive isotopes in the setup. 
	The neutron fluxes and
	energy distributions were taken from published measurements 
	performed in the Gran Sasso. 
	Also simulated was the cosmic 
	muon flux measured in the
	Gran Sasso, on the measured spectrum.
	To give a feeling for the quality of the simulation,
Fig. 
\ref{fig:picUnderTotal} 
	shows the simulated and the measured spectra for a $^{228}$Th 
	source spectrum for 
	as example one of our five detectors. 
	The agreement is excellent.


	The simulation of the background of the experiment 
	reproduces~~  a l l~~ lines observed 
	in the sum spectrum of the five detectors, in the energy range 
	between threshold  (around 100\,keV) and 2020\,keV 
\cite{KK03
}.


\begin{figure}[ht]

\vspace{-0.5cm}
\begin{center}
\includegraphics*[scale=0.45]{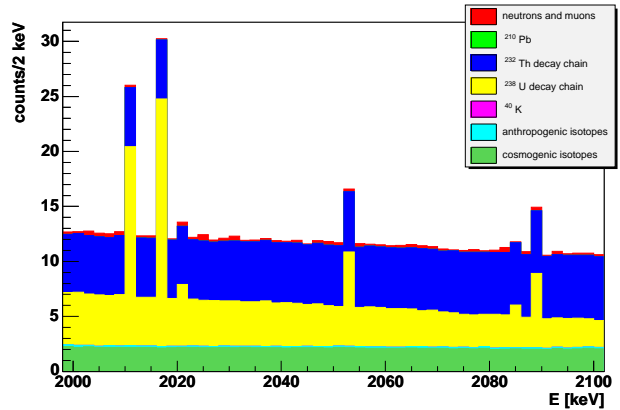}
\end{center}

\vspace{-0.5cm}
\caption{ 
	Simulated background of the
	HEIDELBERG-MOSCOW experiment in the energy range from 2000 
	to 2100\,keV 
	with all known background components, 
	for the period 
	20 November 1995 to 16 April 2002 (from 
\protect\cite{KK03}).
\label{fig:bb_under}}
\end{figure}


	Fig. 
\ref{fig:bb_under} 
	shows the simulated background in the range
	2000-2100\,keV with all~~  
\protect\newline k n o w n~~  background components.

	The background around $Q_{\beta\beta}$ is according 
	to the simulations  f l a t, the
	only expected lines come from $^{214}$Bi (from the $^{238}$U 
	natural decay chain)
	at 2010.89, 2016.7, 2021.6, 2052.94, 2085.1 and 2089.7\,keV. 
	Lines from
	cosmogenically produced $^{56}$Co 
	(at 2034.76\,keV and 2041.16\,keV), 
	half-life 77.3\,days, are not expected since the first 200\,days 
	of measurement of each detector are not used in the data analysis. 
	Also the potential contribution from
	decays of $^{77}$Ge, $^{66}$Ga, or $^{228}$Ac, 
	should not lead to signals visible in our
	measured spectrum near the signal at $Q_{\beta\beta}$. 
	For details we refer to  
\cite{KK03
}.



\vspace{-0.3cm}
\section{Proofs and disproofs}

	Our earlier result published in 
\cite{KK02,KK02-PN,KK02-Found},
	which now is confirmed on a 4$\sigma$ level, had been questioned 
	in some papers 
	[Aalseth et al, hep-ex/0202018, and in 
	Mod. Phys. Lett. A 17 1475-1478; 
	Feruglio et al., Nucl. Phys. B 637 (2002) 345; 
	Zdesenko et al., Phys. Lett. B 546 (2002) 206], and
	Kirpichnikov, talk at Meeting of Physical Section 
	of Russian Academy of Sciences, Moscow, December 2, 2002, 
	(and priv. communication, Dec. 3, 2002) 
	and A. Ianni, Nucl. Instruments A (2004) 
	(available online 28 September 2003). 
	We think that we have shown in a convincing way 
	during 2002 and 2003 that these claims against 
	our results were incorrect in various ways, 
	and have published our arguments in  
\cite{Bi-KK03-NIM,Backgr-KK03-NIM,Support-KK03-PL,KK-StBr02}.
	In particular the estimates of the intensities 
	of the $^{214}{Bi}$ lines in the first two papers 
	do not take into account the effect 
	of true coincidence summing, which can lead to drastic 
	underestimation of the intensities. 
	A correct estimate would also require a Monte Carlo simulation 
	of our setup, which has not been performed in the above papers.

	The paper by Zdesenko et al. starts from an arbitrary assumption, 
	namely that there are lines in the spectrum {\it at best} only 
	at 2010 and 2053\,keV.
	This contradicts to the experimental result, according 
	to which there are further lines in the spectrum 
	(see Fig. 
\ref{fig:Sum90-03}
	in this report). 
	For example they could have easily deduced from the intensity 
	of the 2204\,keV Bi line in the measured spectrum (Fig. 
\ref{fig:Low-HightAll90-03})  
	that lines at 2053\,keV etc. are expected 
\cite{Gromov03}.
	In this way 
	and also by some subtraction procedure, ignoring 
	that the result of subtracting a Poisson-distributed 
	spectrum from a Poisson-distributed spectrum 
	does {\it not} give a Poisson distributed 
	spectrum (see, e.g. 
\cite{NIM99}) 
	they come to wrong conclusions.

	Kirpichnikov states 
\cite{Kir-priv02} 
	that from his analysis he clearly sees the
	2039\,keV line in the full (not pulse-shape discriminated) 
	spectrum on a 4$\sigma$ level.  
	He claims that he does not see the signal in the pulse shape
	spectrum. 
	The simple reason to see less intensity 
	is that in this case he averages
	for determination of the background over 
	the full energy range without
	allowing for any lines.

	All of these papers, when discussing our earlier 
	choice of the width of the search window 
	(in the analysis of the data taken until May 2000), 
	ignore the results of the statistical 
	simulations - we present here, and have published in 
\cite{KK02-PN,KK-antw02,KK-BigArt02,KK02-Found,Support-KK03-PL,KK-StBr02,Backgr-KK03-NIM}.

	The strange effects found recently by the Kurchatov people 
\cite{Kurch03}
	in their rough analysis of part of the data, have been traced back 
	to including corrupt data into the analysis. 
	The artefacts seen in their Figs. 4,5,7,8 
	do not exist in our data, which lead to the results shown 
	in Figs. 
\ref{fig:Low-HightAll90-03},\ref{fig:Sum90-03} 
	(for details see 
\cite{NIM04-NEW-Res,KK04-Big-Repl}).


\vspace{-0.3cm}
\section{Discussion of results}

	We emphasize that we find in all analyses 
	of our spectra a line at the value of Q$_{\beta\beta}$. 
	The results confirm our earlier result with higher 
	statistics. 
	For details we refer to the next Annual Report and to 
\cite{NIM04-NEW-Res}.

	The result obtained is consistent
	 with all other double beta experiments -  
	which reach in general by far less sensitivity. 
	The most sensitive experiments following the 
	HEIDELBERG-MOSCOW experiment are the geochemical $^{128}{Te}$ 
	experiment with 
	${\rm T}_{1/2}^{0\nu} > 2(7.7)\times 10^{24}
	{\rm~ y}$  (68\% c.l.), 
\cite{manuel}
	the $^{136}{Xe}$ experiment by the DAMA group with 
	${\rm T}_{1/2}^{0\nu} > 1.2 \times 10^{24}
	{\rm~ y}$  (90\% c.l.), 
	a second enriched $^{76}{Ge}$ experiment with 
	${\rm T}_{1/2}^{0\nu} > 1.2 \times 10^{24}$ y
\cite{Kirpichn} 
	and a $^{nat}{Ge}$ experiment with 
	${\rm T}_{1/2}^{0\nu} > 1 \times 10^{24}$ y 
\cite{Caldw91}.
	Other experiments are already about a factor of 100 
	less sensitive concerning the \znbb~ 
	half-life: the Gotthard TPC experiment with $^{136}{Xe}$ yields 
\cite{Gottch} 
	${\rm T}_{1/2}^{0\nu} > 4.4 \times 10^{23}
	{\rm~ y}$  (90\% c.l.) and the Milano Mibeta cryodetector experiment 
	${\rm T}_{1/2}^{0\nu} > 1.44 \times 10^{23}
	{\rm~ y}$  (90\% c.l.).

	Another expe\-riment 
\cite{DUM-RES-AVIGN-2000}
	with enriched $^{76}{Ge}$,
	which has stopped operation in 1999 after 
	reaching a significance of 8.8\,kg\,y,
	yields (if one believes their method of 'visual inspection' 
	in their data analysis), in an analysis correcting 
	for on arithmetic error which has been made in 
\cite{DUM-RES-AVIGN-2000}  
	(for discussion see 
\cite{PRD04-CritIGEX})
	a limit of about 
	${\rm T}_{1/2}^{0\nu} > 5 \times 10^{24}
	{\rm~ y}$  (90\% c.l.). 
	The $^{128}{Te}$ geochemical experiment 
	yields $\langle m_\nu \rangle < 1.1$ eV (68 $\%$ c.l.)
\cite{manuel},   
	the DAMA $^{136}{Xe}$ experiment 
	$\langle m_\nu \rangle < (1.1-2.9)$\,eV 
	and the $^{130}{Te}$ cryogenic experiment yields 
	$\langle m_\nu \rangle < 1.8$\,eV. 

	Concluding we obtain, with $>$ 4$\sigma$ probability, 
	evidence for a neutrinoless 
	double beta decay signal. 
	Following this interpretation, at this confidence level, 
	lepton number is not conserved. 
	Further the neutrino is a Majorana particle. 
	If the 0$\nu\beta\beta$ amplitude is dominated by exchange 
	of a massive neutrino the effective mass 
	$\langle m \rangle $ is deduced 
	from the full spectrum (using the matrix elements of 
\cite{Sta90})
	to be 
	$\langle m \rangle$ = (0.1 - 0.9)\,eV (3$\sigma$ confidence range), 
	allowing already for a $\pm$ 50\% uncertainty 
	of the matrix element. The best value is 0.4\,eV.

	Assuming other mechanisms to dominate the \znbb~ decay amplitude, 
	the result allows to set stringent limits on parameters of SUSY 
	models, leptoquarks, compositeness, masses of heavy neutrinos, 
	the right-handed W boson and possible violation of Lorentz 
	invariance and equivalence principle in the neutrino sector. 
	For a discussion and for references we refer to 
\cite{KK60Y,KK-Bey97,KK-Neutr98,KK-SprTracts00,KK-NANPino00,KKS-INSA02}.

\begin{figure}[t]
\begin{center}
\includegraphics[scale=0.5]{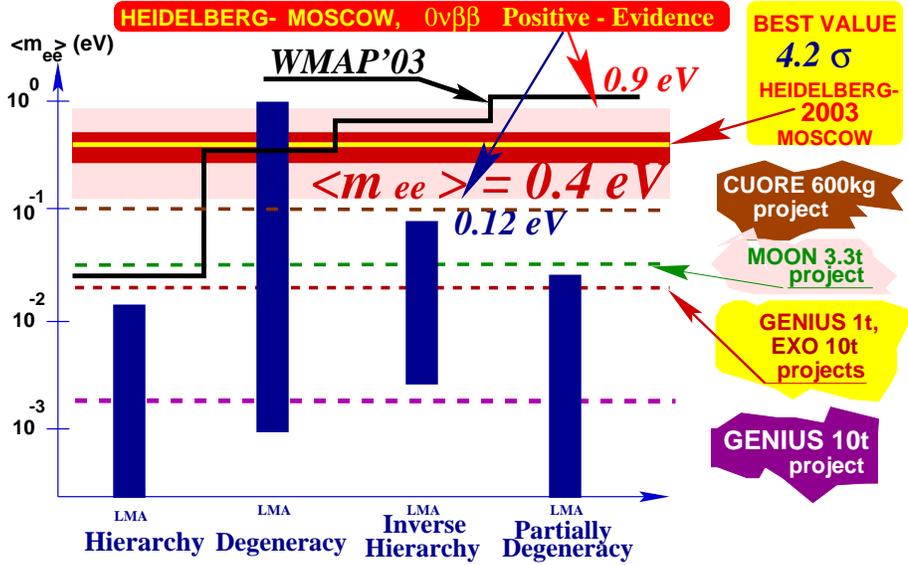}
\end{center}
\caption{
	The impact of the evidence obtained at (4.2$\sigma$ c.l.) 
	for neutrinoless double beta decay (best value 
	of the effective neutrino mass 
	$\langle m \rangle$ = 0.4\,eV, 3$\sigma$ 
	confidence range (0.1 - 0.9)\,eV - 
	allowing already for an uncertainty of the nuclear 
	matrix element of a factor of $\pm$ 50$\%$) 
	on possible neutrino mass schemes. 
	The bars denote allowed ranges of $\langle m \rangle$ 
	in different neutrino mass scenarios, 
	still allowed by neutrino oscillation experiments (see 
\protect\cite{KK-Sark01,KK-S03-WMAP}). 
	All models except the degenerate one are excluded by the 
	new \znbb ~~decay result. Also shown is the exclusion line from WMAP, 
	plotted for $\sum m_{\nu} < 1.0\, eV$ 
\protect\cite{Hannes03}. 
	WMAP does not rule out any of the neutrino mass schemes. 
	Further shown are the expected sensitivities 
	for the future potential double beta experiments 
	CUORE, MOON, EXO  
	and the 1 ton and 10 ton project of GENIUS 
\protect\cite{KK60Y,KK-SprTracts00,KK-00-NOON-NOW-NANP-Bey97-GEN-prop,GEN-prop} 
	(from 
\protect\cite{KK-S03-WMAP}).
\label{fig:Jahr00-Sum-difSchemNeutr}}
\end{figure}

	With the value deduced for the effective neutrino mass,  
	the HEIDELBERG-MOSCOW experiment excludes several 
	of the neutrino mass scenarios 
	allowed from present neutrino oscillation experiments
	(see Fig.
\ref{fig:Jahr00-Sum-difSchemNeutr}) 
	- allowing only for a degenerate mass scenario  
\cite{KK-Sark01,KK-S03-WMAP,NIM04-NEW-Res}. 
	Fig. 
\ref{fig:Jahr00-Sum-difSchemNeutr}
	shows also the limits obtained from WMAP, 
	which at the present level of sensitivity 
	is not able to rule out any neutrino mass scheme.


	The evidence for neutrinoless double beta decay has been 
	supported by various recent experimental 
	and theoretical results (see Table 1).
	Assuming the degenerate scenarios to be realized in nature 
	we fix - according to the formulae derived in 
\cite{KKPS} - 
	the common mass eigenvalue of the degenerate neutrinos 
	to m = (0.1 - 3.6)\,eV. 
	Part of the upper range is excluded by 
	tritium experiments, which give a limit 
	of m $<$ (2.2 - 2.8)\,eV (95$\%$ c.l.) 
\cite{Weinh-Neu00
}.
	The full range can only  partly 
	(down to $\sim$ 0.5\,eV) be checked by future  
	tritium decay experiments,  
	but might be checked by some future $\beta\beta$ 
	experiments (see next section). 
	Recent theoretical work 
\cite{Kirch04} 
	even doubts, that tritium experiments  
	are in {\it principle} capable to check a \znbb result.
	The deduced best value for the mass 
	is consistent with expectations from experimental 
	$\mu ~\to~ e\gamma$
	branching limits in models assuming the generating 
	mechanism for the 
	neutrino mass to be also responsible 
	for the recent indication for as anomalous magnetic moment 
	of the muon
\cite{MaRaid01}.  
	It lies in a range of interest also for Z-burst models recently 
	discussed as explanation for super-high energy cosmic ray events 
	beyond the GKZ-cutoff 
\cite{Farj00-04keV,
FKR01} 
	and requiring neutrino masses in the range (0.08 - 1.3)\, eV. 
	A recent model with underlying A$_4$ symmetry for 
	the neutrino mixing matrix also leads to degenerate Majorana 
	neutrino masses $>$ 0.2\, eV, consistent with the present result 
	from \znbb~ decay 
\cite{Ma-DARK02,BMV02}. 
	The result is further consistent with the theoretical paper of 
\cite{Moh03}.
	Starting with the hypothesis that quark and lepton mixing are
	identical at or near the GUT scale, Mohapatra \etal 
~\cite{Moh03} show
	that the large solar and atmospheric neutrino mixing angles can be
	understood purely as result of renormalization group evolution, if
	neutrino masses are quasi-degenerate (with same CP parity). 
	The common 
	Majorana neutrino mass then must be, in this model, 
	larger than 0.1 eV. 
	An completely independent theoretical proof, 
	that neutrinos should have Majorana nature, 
	has been given recently by 
\cite{Hof04}.

	For WMAP a limit for the sum of the  neutrino masses of 
	$m_s = \sum m_i < 0.69 ~{\rm eV~ at~ } 95\%$ c.l.,
	was given by the analysis of ref. 
\cite{WMAP03}. 
	More realistically this limit on the total 
	mass should be 
\cite{Hannes03}
$m_s = \sum m_i < 1.0 ~{\rm\,eV\, at }~ 95\% {\rm c.l.}$ 
	The latter analysis also shows, that four 
	generations of neutrinos are still allowed and in the
	case of four generations the limit on the total mass is 
	increased to $1.38$ eV. If there is
	a fourth neutrino with very small mass, then the limit 
	on the total mass of the three neutrinos is even further weakened 
	and there is essentially no
	constraint on the neutrino masses. 
	In our Fig. 
\ref{fig:Jahr00-Sum-difSchemNeutr}
	we show the contour line for WMAP assuming 
	$\sum m_i < 1.0 ~{\rm eV}$.
 	
	A recent analysis of the Sloan Digital Sky Survey, 
	together with WMAP yields 
\cite{Barger03}
\vspace{-0.3cm}
\begin{equation}
m_s = \sum m_i < 1.7 ~{\rm eV~~~~~~~ at }~ 2\sigma. 
\end{equation}
	Comparison of the WMAP results with the effective mass 
	from double beta decay rules out completely (see 
\cite{Muray03})
	a 15\,years old old-fashioned 
	nuclear matrix element of double beta decay, used
	in a recent analysis of WMAP  
\cite{Vogel1}. 
	In that calculation of the nuclear matrix element there was 
	not included a realistic nucleon-nucleon
	interaction, which has been included by all other calculations of the
	nuclear matrix elements over the last 15\,years.

	The range of $\langle m \rangle $ fixed in this 
	work is, in the range to be explored 
	by the satellite experiments MAP and PLANCK 
\cite{
KKPS,WMAP03,Hannes03}. 
	The limitations of the information from WMAP are seen in Fig. 
\ref{fig:Jahr00-Sum-difSchemNeutr}, 
	thus results of PLANCK are eagerly awaited.

	The neutrino mass deduced leads to 0.002$\ge \Omega_\nu h^2 \le$
	0.1 and thus may allow neutrinos to still play 
	an important role as hot dark matter in the Universe 
\cite{KK-LP01}.


\vspace{-0.3cm}
\section{Future of $\beta\beta$ experiments}

	With the HEIDELBERG-MOSCOW experiment, the era of the small smart 
	experiments is over. 
	New approaches and considerably {\it enlarged experiments} 
	(as discussed, e.g. in 
\cite{KK-InJModPh98,KK-Bey97,KK60Y,KK-Neutr98,KK-00-NOON-NOW-NANP-Bey97-GEN-prop,GEN-prop,KK-NOW00,KK-LP01})
	will be required in future 
	to fix the  \znbb half life of $^{76}{Ge}$ with higher accuracy. 
	This will, however, because of the uncertainties 
	in the nuclear matrix elements, which probably hardly 
	{\it can} be reduced to less than 50\%, 
	{\it only marginally reduce} 
	the precision {\it of the deduced neutrino mass}.
	
	Since it was realized in the HEIDELBERG-MOSCOW experiment, 
	that the remaining small background is coming from the material 
	close to the detector (holder, copper cap, ...), 
	elimination of {\it any} material close to the detector 
	will be decisive. Experiments which do not take this 
	into account, 
	will allow at best only rather limited steps in sensitivity. 
	Furthermore there is the problem in cryodetectors that they 
	{\it cannot} differentiate between a $\beta$ and a $\gamma$ signal, 
	as this is possible in Ge experiments.

	Another crucial point is the energy resolution, 
	which can be optimized {\it only} in experiments 
	using Germanium detectors, or, to some less extent, with bolometers. 
	It will be difficult to probe evidence for this rare decay 
	mode in experiments, which have to work - as result of their 
	limited resolution - with energy windows around 
	Q$_{\beta\beta}$ of several hundreds of keV.

	Another important point is the efficiency 
	of a detector for detection of a $\beta\beta$ signal.
	For example, with 14$\%$ efficiency a potential 
	future 100\,kg $^{82}{Se}$ 
	experiment would be, because 
	of its low efficiency, equivalent only to a 10\,kg 
	experiment (not talking about the energy resolution).

	In the first proposal for a third generation double 
	beta experiment, our GENIUS proposal 
\cite{KK-Bey97,KK-InJModPh98,KK-H-H-97,KK-Neutr98,KK-00-NOON-NOW-NANP-Bey97-GEN-prop,GEN-prop},
	the idea is to use 'naked' Germanium detectors in a huge tank 
	of liquid nitrogen. It seems to be at present the {\it only} 
	proposal, which can fulfill {\it both} requirements 
	mentioned above - to increase the detector mass 
	and simultaneously reduce the background drastically. 
	At the  {\it present} status of results of 
	the \HM experiment, 
	however - with a confidence level of $\sim 4 \sigma$, 
	it is {\it questionable}, whether GENIUS 
	would be needed for $\beta\beta$ decay.
	Probably it would be preferable to perform 
	an experiment with another isotope {\it but fulfilling}  
	all requirements mentioned above. 
	The GENIUS-Test-Facility, originally planned 
	to prove the feasibility of some key constructional parameters 
	of GENIUS, and put into operation on May 5, 2003 in GRAN SASSO, 
	could however, play an important role in testing the evidence seen 
\cite{DAMA-03} 
	for cold dark matter by DAMA (see 
\cite{TF-NIM03,GenTF-0012022}, 
	and another Report to this volume).  
	Only a GENIUS with some ten tons of enriched $^{76}{Ge}$ 
	might possibly be of interest,  
	to investigate whether another exotic mechanism 
	such as exchange of SUSY particles, 
	(see, e.g. 
\cite{KK60Y})
	might contribute to the \znbb~ decay amplitude. 
	This may be, however, a very far dream. 



\vspace{-0.5cm}
\section{Summary}
	
	The \HM experiment has been continued regularly in 2003.  
	Unfortunately, it had to stop operation according 
	to non-prolongation of our contract with Kurchatov institute,  
	at 30 November 2003. 
	Since then still various calibration measurements 
	with radioactive sources are going on.

	The first analysis of the full data taken with 
	the \HM experiment in the period 2 August 1990 
	until 20 May 2003 is presented. 
	The improved statistics and data analysis leads  
	to a $\sim 4 \sigma$ evidence for a signal 
	at the Q-value for neutrinoless double beta decay. 
	This confirms our earlier claim 
\cite{KK02,KK02-PN,KK02-Found,NIM04-NEW-Res}. 	
	Additional support for this evidence has been presented  by
	showing consistency of the result - for the signal,  
\mbox{a n d}  
	for the background - with other double beta decay 
	experiments using non-enriched or enriched Germanium detectors 
	(see also 
\cite{Support-KK03-PL,Backgr-KK03-NIM}). 
	In particular it has been
	shown that the lines seen in the vicinity of the signal 
	are seen also in the other experiments. This is important 
	for the correct treatment of the
	background. Furthermore, the sensitivity of the peak identification
	procedures has been demonstrated 
	by extensive statistical simulations.
	It has been further shown by new extensive simulations 
	of the expected
	background by GEANT4, that the background around 
	$Q_{\beta\beta}$ should be flat, and
	that no known gamma line is expected at the energy 
	of $Q_{\beta\beta}$ (see 
\cite{KK03}).
	The 2039\,keV signal is seen  \mbox{o n l y}  in
	the HEIDELBERG-MOSCOW  experiment, which has a {\it by far larger} 
	statistics than all other double beta experiments.

	The importance of first evidence for violation 
	of lepton number and of the Majorana nature of neutrinos 
	is obvious. It requires beyond Standard Model Physics 
	on one side, and 
	may open a new era in space-time structure 
\cite{AHLUW02}.
	It has been discussed that the Majorana nature of the neutrino 
	tells us that spacetime does realize a construct 
	that is central to construction of supersymmetric theories.

	One of the consequences of the result of the \HM experiment 
	on the present confidence level, 
	may be, that to obtain {\it deeper} information 
	on the process of neutrinoless double beta decay, 
	{\it new experimental approaches, different from all}, 
	{\it what is at present persued}, may be required. 
	The unique importance of double beta decay to  
	investigate the neutrino mass, 
	is stressed by the recent observation, 
	that tritium experiments might suffer from principle problems  
	to see a neutrino mass at all 
\cite{Kirch04}.

	With the successful start of operation of GENIUS-TF 
	with the first four naked Ge detectors in liquid nitrogen 
	on May 5, 2003 in GRAN SASSO, which is described in 
\cite{CERN03-GenTF,TF-NIM03} 
	(see our second contribution to this Report) 
	a historical step has been achieved of a novel technique and into 
	a new domain of background reduction in underground 
	physics in the search for rare events. 
	In the light of the above comments, natural task 
	of GENIUS-TF will be to look for cold dark matter 
	by the modulation signal.


\vspace{0.3cm}
\noindent
{\large\bf Acknowledgement:}

	The authors would like to thank all colleagues, 
	who have contributed to the experiment.
	Our thanks extend also to the technical staff of the 
	Max-Planck Institut f\"ur Kernphysik and 
	of the Gran Sasso Underground Laboratory.  
	We acknowledge the invaluable support from BMBF 
	and DFG, and LNGS of this project.
	We are grateful to the former State Committee of Atomic 
	Energy of the USSR for providing the monocristalline Ge shielding 
	material used in this experiment.


\vspace{0.3cm}
\noindent
{\Large\bf List of Edited Proceedings (2003)}                   

\begin{enumerate}

\item 
	H.V. Klapdor-Kleingrothaus (ed.)
	{\sf Physics Beyond the Standard Model: Beyond the Desert 02}, 
	Proc. of Intern. Conf. BEYOND'02, Oulu, Finland, 2-7 Jun 2002, 
	IOP, Bristol, 2003, 734 pages.

\item
	H.V. Klapdor-Kleingrothaus (ed.)
	{\sf Physics Beyond the Standard Model: Beyond the Desert 03}, 
	Proc. of Intern. Conf. BEYOND'03, Tegernsee, Germany, 4-9 June 2003, 
	Springer, Heidelberg, 2004 (in preparation).

\end{enumerate}

\noindent
{\Large\bf List of Publications (2003)} 

\begin{enumerate}

\item 
	H.V. Klapdor-Kleingrothaus, A. Dietz, 
	I.V. Krivosheina, Ch. D\"orr and C. Tomei, 
	Phys. Lett. B 578 (2004) 54-62 and hep-ph/0312171, 
	{\it ``Support of Evidence for Neutrinoless Double Beta Decay''.}

\item 
	A. D\"orr and H.V. Klapdor-Kleingrothaus, 
	Nucl. Instrum. Meth. A 513 (2003) 596-621, 
	{\it ``New Monte-Carlo simulation of the \HM double beta 
	decay experiment''.}

\item 
	 H.V. Klapdor-Kleingrothaus, O. Chkvorez, 
	I.V. Krivosheina, C. Tomei ,
	Nucl. Instrum. Meth. A 511 (2003) 335-340 and hep-ph/0309157, 
	{\it ``Measurement of the $^{214}{Bi}$ spectrum in the energy region 
	around the Q-value of $^{76}{Ge}$ neutrinoless double-beta decay''.}

\item 
	H.V. Klapdor-Kleingrothaus, A. Dietz, I.V. Krivosheina, 
	C. D\"orr, C. Tomei, 
	Nucl. Instrum. Meth. A 510 (2003) 281-289 and hep-ph/0308275, 
	{\it ``Background Analysis around $Q_{\beta\beta}$ for $^{76}{Ge}$ 
	Double Beta Decay experiments, and Statistics at Low Count Rates''.}

\item 
	 H.V. Klapdor-Kleingrothaus and V. Bednyakov, 
	CERN Courier 43 N2 (2003) 29-30, 
	{\it ``Neutrinos Lead Beyond the Desert''.}

\item 
	 H.V. Klapdor-Kleingrothaus and U. Sarkar,
	Mod. Phys. Lett. A 18 (2003) 2243 
	\& hep-ph/0304032, 
	{\it ``Consequences of neutrinoless double beta decay and WMAP''.}

\item 
	H.V. Klapdor-Kleingrothaus, 
	 Int. J. Mod. Phys. A 18 (2003) 4113 
	\& hep-ph/0303217, 
	{\it ``To be or not to Be? - 
	First Evidence for Neutrinoless Double Beta Decay''.}

\item 
	G. Bhattacharyya, H.V. Klapdor-Kleingrothaus, 
	H. P\"as, A. Pilaftsis,  
	Phys. Rev. D 67 (2003) 113001 and hep-ph/0212169,
	{\it ``Neutrinoless Double Beta Decay from Singlet Neutrinos 
	in Extra Dimensions''.}

\item
	H.V. Klapdor-Kleingrothaus, Utpal Sarkar, 
	Phys. Lett. B 554 (2003) 45-50 and hep-ph/0211274, 
	{\it ``Neutrinoless double beta decay with scalar bilinears''.} 

\item 
	H.V. Klapdor-Kleingrothaus, A. Dietz, G. Heusser, 
	I.V. Krivosheina, D. Mazza, H. Strecker, C. Tomei, 
	Astropart.Phys. 18 (2003) 525-530 and hep-ph/0206151,
	{\it ``First Results from the HDMS experiment in the Final Setup''.}

\item 
	H.V. Klapdor-Kleingrothaus, 
	hep-ph/0302237 and in Indian National 
	Science Academy (INSA) - Special issue: Neutrinos, 
	(2003) eds. D. Indumathi, G. Rajasekaran and M.V.N. Murthy; 
	{\it ``One Year of Evidence for Neutrinoless Double Beta Decay''.}

\item
	H.V. Klapdor-Kleingrothaus,  
	Zacatecas Forum in Physics 2002, Zacatecas, 
	Mexico, 11-13 May 2002, 
	Found. Phys. 33 (2003) 813-829, and hep-ph/0302234, 
	{\it ``First Evidence for Neutrinoless Double Beta Decay''.}

\end{enumerate}


\noindent
{\Large\bf List of Contributions to Conferences (2003)}

\begin{enumerate}

\item
	H.V. Klapdor-Kleingrothaus and I.V. Krivosheina, in 
	Proc. of Intern. Conf. IHEPP03, Valencia, September 2003, 
	PRHEP-AHEP2003/060, 
	{\it ``Status of Absolute Neutrino Mass and Double Beta Decay''.}

\item
	G. Bhattacharyya, H.V. Klapdor-Kleingrothaus, H. P\"as$^*$, 
	A. Pilaftsis, 
	 in Proc. of Intern. Conf. IHEPP03, Valencia, September 2003, 
	{\it ``Double beta decay and the extra-dimensionsinal 
	seesaw mechanism''.}

\item
	H.V. Klapdor-Kleingrothaus, in 
	Proc. of Intern. Conf. BEYOND'03, 
	{\sf Physics Beyond the Standard Model: Beyond the Desert 03}, 
	Tegernsee, Germany, 4-9 June 2003, 
	Springer, Heidelberg, 2004, H.V. Klapdor-Kleingrothaus (ed.),  
	{\it "First Evidence for Neutrinoless Double Beta Decay - and World
                    Status of the Absolute Neutrino Mass".}

\item
	H.V. Klapdor-Kleingrothaus, in 
	Proc. of Arbeitstreffen {\sf "Hadronen und Kerne"}, 
	Mei\ss{}en, St.-Afra-Klosterhof, Germany, 8.-11. September 2003, 
	{\it "Aktueller Status des Neutrinolosen Doppel-Beta-Zerfalls".}

\item
	H.V. Klapdor-Kleingrothaus in 
	Proc. of Intern. Conf. BEYOND'02, 
	{\sf Physics Beyond the Standard Model: Beyond the Desert 02}, 
	H.V. Klapdor-Kleingrothaus (ed.),  
	Oulu, Finland, 2-7 Jun 2002, IOP, Bristol, 2003, 215 - 240,
	{\it ``Further support of evidence for neutrinoless 
 	double beta decay''.}

\item
	H.V. Klapdor-Kleingrothaus and U. Sarkar, in  
	Proc. of Intern. Conf. BEYOND'02, 
	{\sf Physics Beyond the Standard Model: Beyond the Desert 02}, 
	H.V. Klapdor-Kleingrothaus (ed.),
	Oulu, Finland, 2-7 Jun 2002, IOP, Bristol, 2003, 253 - 269,
	{\it ``Consequences of neutrinoless double beta decay''.}

\item 
	H.V. Klapdor-Kleingrothaus$^*$, A. Dietz, I.V. Krivosheina, 
	Nucl. Phys. Proc. Suppl. 124: 209-213, 2003, D. Cline (ed.), 
	for 5th Intern. UCLA Symposium on Sources 
	and Detection of Dark Matter and Dark Energy in the Universe (DM
	2002), Marina del Rey, California, 20-22 Feb 2002, 
	{\it ``Search for Cold and Hot Dark Matter 
	with the HEIDELBERG-MOSCOW Experiment, HDMS, GENIUS and GENIUS-TF''.}

\item 
	H.V. Klapdor-Kleingrothaus, in Proc. of 
        International Europhysics Conference on High Energy Physics, 
	European Physical Society (EPS), Aachen, Germany, 17 - 23 July 2003
	{\it "The GENIUS Test Facility in GRAN SASSO".}

\item 
	H.V. Klapdor-Kleingrothaus, 
	Deutsche Physika\-lische Gesell\-schaft e. V. (DPG), 
        Aachen, Germany, 24.-28. M\"arz, 2003, 
      	{\it"Status of Evidence for Neutrinoless Double Beta Decay"}. 

\item 
	Ch. D\"orr$^*$ and H.V. Klapdor-Kleingrothaus, 
	Deutsche Physika\-lische Gesell\-schaft e. V. (DPG), 
        Aachen, Germany, 24.-28. M\"arz, 2003, 
   	{\it "Neue Ergebnisse f\"ur den neutrinobegleiteten Doppelbetazerfall
        von $^{76}{Ge}$ im Rahmen des HEIDELBERG-MOSCOW Experiments"}.

\item
	H.V. Klapdor-Kleingrothaus, in 
	Proc. of Intern. Conf. SUGRA2003, 
	{\sf ``20 years of SUGRA and the Search for SUSY and Unification''}, 
	Boston, USA, March 17-20, 2003, 
	{\it ``Status of Evidence for Neutrinoless Double Beta Decay 
	from the \HM Experiment - and Implications for Supersymmetry''}.

\item 
	H.V. Klapdor-Kleingrothaus, 
	in Proc. of Neutrinos and implications for physics beyond 
	the standard model, 
	Stony Brook, USA, 11-13 October, 2002, 4113-4128,  
	 Int. J. Mod. Phys. A 18 (2003) 4113-4128 and hep-ph/0303217, 
	{\it ``To be or not to Be? - 
	First Evidence for Neutrinoless Double Beta Decay''}.

\item 
	H. V. Klapdor-Kleingrothaus, 
	in Proc. of NOON 2003, Japan, Kanazawa, February 2003, 
	World Scientific 2004, eds. Y. Suzuki, hep-ph/0307330, 
	{\it ``Status of Evidence for Neutrinoless Double Beta Decay, 
	and the Future: GENIUS and GENIUS-TF.''}

\end{enumerate}


\noindent
{\Large\bf List of Colloquia and Seminars Made During 2003}

\begin{enumerate}

\item H.V. Klapdor-Kleingrothaus, 
	Physikalisches Institut, Theorie Department, Bonn, 7. November 2003,
	{\it "Absolute Neutrino Mass After the First  
	Evidence for Neutrinoless Double Beta Decay - 
	and Implications of Double Beta Decay for Exotic Physics".}

\item H.V. Klapdor-Kleingrothaus,
	DESY, Zeuthen, Germany, 17 Juli (2003),
	{\it "First Evidence for Neutrinoless Double Beta Decay - 
         and World Status of the absolute Neutrino Mass."}

\item H.V. Klapdor-Kleingrothaus, 
	 Fachbereich Physik, Bergische Universit\"at Wuppertal, Germany 
                  26 May 2003, 
	{\it "The Absolute Neutrino Mass After the First  
	Evidence for Neutrinoless Double Beta Decay".}

\item H.V. Klapdor-Kleingrothaus, 
	Physikalisches Institut, Fakult\"at f\"ur Physik und Astronomie,
                Ruprecht-Karls-Universit\"at, Heidelberg, Germany
                          28 April 2003, 
	{\it "Status of Evidence for Neutrinoless Double Beta Decay"}.

\item H.V. Klapdor-Kleingrothaus,
	 Osaka University, Japan, 
                          17 February 2003,
	{\it "First Evidence for Neutrinoless Double Beta Decay
               from the HEIDELBERG-MOSCOW Experiment 
        and Implication for Particle Physics and Astrophysics".}

\item H.V. Klapdor-Kleingrothaus, 
	Institut f\"ur Physik, Universit\"at Mainz, Germany, 
                          22 Januar 2003, 
		{\it "First Evidence for Neutrinoless Double Beta Decay
                        - and Future of the Field".}
\end{enumerate}



\vspace{-0.5cm}
        


\begin{thebibliography}{99}


\bibitem{KK02}
	H.V. Klapdor-Kleingrothaus et al.
	{\it Mod. Phys. Lett.} 
	{\bf A 16} (2001) 2409 - 2420.

\bibitem{KK02-PN}
	H.V. Klapdor-Kleingrothaus, A. Dietz, I.V. Krivosheina,  
	Part. \& Nucl. 110(2002)57.

\bibitem{KK-antw02}
	H.V. Klapdor-Kleingrothaus, hep-ph/0205228, in 
	Proc. of DARK2002, 
	Cape Town, South Africa, February 4 - 9, 2002,
	eds. by 
	H.V. Klapdor-Kleingrothaus and R.D. Viollier,  
	Springer (2002) 404 - 411.

\bibitem{KK-BigArt02}
	H.V. Klapdor-Kleingrothaus, hep-ph/0302248, Proc.DARK2002, 
	Cape Town, South Africa, February 4 - 9, 2002, 
	eds. by 
	H.V. Klapdor-Kleingrothaus and R.D. Viollier,  
	Springer (2002) 367 -- 403.

\bibitem{KK02-Found}
	H.V. Klapdor-Kleingrothaus, A. Dietz and I.V. Krivosheina,  
	{\it Foundations of Physics} {\bf 31} (2002) 1181-1223  
	and Corr., 2003: 
	{\footnotesize  http://www.mpi-hd.mpg.de/non\_acc/main\_results.html}.

\bibitem{NIM04-NEW-Res}
	H.V. Klapdor-Kleingrothaus, et al., {\it Nucl. Instr. Meth.} 
	{\bf 522 A} (2004) 371-406 and hep-ph/0403018 and 
	{\it Phys. Lett.} {\bf B 586} (2004) 198-212.

\bibitem{Support-KK03-PL}
	H.V. Klapdor-Kleingrothaus,  A. Dietz, I.V. Krivosheina, 
	Ch. D\"orr, C. Tomei, 
	{\it Phys. Lett.} {\bf B 578} (2004) 54-62 and hep-ph/0312171.

\bibitem{HDM01}
	H.V. Klapdor-Kleingrothaus et al., (HEIDELBERG-MOSCOW Col.), 
	{\it Eur. Phys. J.} {\bf A 12} (2001) 147,
	Proc. of "Third Intern. Conf. on Dark Matter
	in Astro- and Particle Physics", DARK2000, ed. 
	H.V. Klapdor-\-Kleingrothaus, 
	Springer (2001) 520 - 533. 


\bibitem{replay} 
	H.V. Klapdor-Kleingrothaus, 
	hep-ph/0205228, and in Proc. of DARK2002, 
	Cape Town, South Africa, February 4 - 9, 2002, eds. by 
	H.V. Klapdor-Kleingrothaus and R.D. Viollier,  
	Springer, Heidelberg (2002) 404 - 411.


\bibitem{KK-StBr02}
	H.V. Klapdor-Kleingrothaus, hep-ph/0303217 and 
	in Proc. of ``Neutrinos and 
	Implications for Phys. Beyond the SM'', 
	Stony Brook, 11-13 Oct. 2002, World Scientific (2003) pp. 367-382.


\bibitem{Backgr-KK03-NIM}
	H.V. Klapdor-Kleingrothaus,  A. Dietz, I.V. Krivosheina, 
	Ch. D\"orr, C. Tomei, {\it Nucl. Instr. Meth.}
	{\bf 510 A} (2003) 281-289 and hep-ph/0308275. 


\bibitem{Bi-KK03-NIM}
	H.V. Klapdor-Kleingrothaus,  O. Chkvorez, I. V. Krivosheina,  
	C. Tomei, {\it Nucl. Instr. Meth.} 
	{\bf 511 A} (2003) 335-340 and hep-ph/0309157.



\bibitem{KK03}
	Ch. D\"orr and H.V. Klapdor-Kleingrothaus, 
	{\it Nucl. Instr. Meth.} 
	{\bf 513 A} (2003) 596-621.


\bibitem{KKPS}
	H.V. Klapdor-Kleingrothaus, H. P\"as and A.Yu. Smirnov, 
	{\it Phys. Rev.} {\bf D 63} (2001) 073005 and 
	{\it hep-ph/}{\bf 0003219}.
%

\bibitem{KK-Sark01}
	H.V. Klapdor-Kleingrothaus and U. Sarkar, 
	{\it Mod. Phys. Lett.} {\bf A 16} (2001) 2469 - 2482.

\bibitem{KKS-INSA02}
	H V Klapdor-Kleingrothaus, 
	Special issue: Neutrinos, 2003, Proc. Indian Natl. Sci. Acad.,
	 hep-ph/0302237.

\bibitem{KK-InJModPh98}
	H.V. Klapdor-Kleingrothaus, {\it Int. J. Mod. Phys.} {\bf A 13}   
	(1998) 3953.

\bibitem{KK-SprTracts00}
	H.V. Klapdor-Kleingrothaus, {\it Springer Tracts in Modern Physics}, 
	{\bf 163} (2000) 69 - 104, 
	{\it Springer-Verlag, Heidelberg, Germany} (2000).

\bibitem{KK60Y}
		H.V. Klapdor-Kleingrothaus, 
		{\sf "60 Years of Double Beta Decay - From
	Nuclear Physics to Beyond the Standard Model"}, 
	{\it World Scientific, 
		Singapore} (2001) 1281~p.

\bibitem{KK-S03-WMAP}
	H.V. Klapdor-Kleingrothaus and U. Sarkar, 
	hep-ph/0304032, and  
	{\it Mod. Phys. Lett.} {\bf A 18} (2003) 2243-2254.

\bibitem{KK-Cp-parity03}
	H.V. Klapdor-Kleingrothaus, to be publ. 2004,  
	and in Proc. of  
	Third Intern. Conf. on Particle Physics 
	Beyond the Standard Model, BEYOND02, Oulu, 
	Finland, 2-7 June 2002, ed. by H.V. Klapdor-Kleingrothaus, 
	IOP, Bristol 2003, 215 - 240.

\bibitem{HK-KK97-98}
	M. Hirsch, H.V. Klapdor-Kleingrothaus and S.G. Kovalenko, 
	{\it Phys. Lett.} {\bf B 398} (1997) 311, 
	{\it Phys. Lett.} {\bf B 403} (1997) 291, and 
	{\it Phys. Rev.} {\bf D 57} (1998) 1947.

\bibitem{HK-KK95-99}
	M. Hirsch, H.V. Klapdor-Kleingrothaus and S.G. Kovalenko, 
	{\it Phys. Rev. Lett.} {\bf 75} (1995) 17, and 
	{\it Phys. Rev.} {\bf D 53} (1996) 1329, and 
	{\it Phys. Lett.} {\bf B 372} (1996) 181; and 
	G. Bhattacharya,  H.V. Klapdor-Kleingrothaus, H. P\"as 
	{\it Phys. Lett.} {\bf B 463} (1997) 77.
 

\bibitem{Random}
        CERN number generators (see e.g. 
	{\footnotesize http://root.cern.ch/root/html/TRandom.html})


\bibitem{Sta90}
	A. Staudt, K. Muto and H.V. Klapdor-Kleingrothaus,  
	{\it Eur. Lett.} {\bf 13} (1990) 31.



\bibitem{Tabl-Isot96}
	R.B. Firestone and V.S. Shirley, 
	Table of Isotopes, 8th Ed., 
	{\it John W.\%S}, N.Y. (1998).

\bibitem{gamma}
        G. Gilmore et al.
        ``Practical Gamma-Ray Spectr.'', 
        Wiley and Sons (1995).

\bibitem{New-Q-2001}
	G. Douysset et al.,
	{\it Phys. Rev. Lett.} {\bf 86} (2001) 4259-4262.

\bibitem{Old-Q-val}
	J.G. Hykawy et al., 
	{\it Phys. Rev. Lett.} {\bf 67} (1991) 1708.

\bibitem{Q-val-Audi}
	G. Audi, A.H. Wapstra, 
	Nucl. Phys. A 595 (1995) 409-480.

\bibitem{Q-val-Ellis}
	R.J. Ellis et al., Nucl. Phys. 
	A 435 (1985) 34-42.




\bibitem{Gromov03}
	K. Ya. Gromov, priv. communication, 2003.

\bibitem{NIM99}
	M.D. Hannam, W.J. Thompson,  
	Nucl. Instr. Meth. A 431 (1999) 239-251.



\bibitem{Caldw91}
	D. Caldwell,
	{\it J. Phys. G }{\bf 17}, S137-S144 (1991).

\bibitem{Kirpichn}
	I.V. Kirpichnikov et al. 
	Mod. Phys. Lett. A 5 (1990) 1299 
	Preprint ITEP, 1991. 

\bibitem{Kir-priv02}
	I.V. Kirpichnikov, priv. communication, Dec. 3, 2002.


\bibitem{manuel}
	O. Manuel et al., 
	in Proc. Intern. Conf. Nuclear Beta Decays 
	and the Neutrino, eds. T. Kotani et al., 
	World Scientific (1986) 71, 
	{\it J. Phys. G: Nucl. Part. Phys.} {\bf 17} (1991) S221-S229; 
	T. Bernatovicz et al. {\it Phys. Rev. Lett.} {\bf 69} 
	(1992) 2341-2344.


\bibitem{Gottch}
	R. L\"uscher et al., {\it Phys. Lett.} (1998) 407.


\bibitem{KK-Bey97}
	H.V. Klapdor-Kleingrothaus in Proc. of BEYOND'97, 
	First Intern. Conf. on Particle Physics 
	Beyond the Standard Model, Castle Ringberg, 
	Germany, 8-14 June 1997, 
	ed. by H.V. Klapdor-Kleingrothaus and H. P\"as, 
	{\it IOP Bristol} (1998) 485-531.  

\bibitem{KK-H-H-97}
	H.V. Klapdor-Kleingrothaus, J. Hellmig \& M. Hirsch, 
	{\it J. Phys.} {\bf G 24} (1998) 483-516.

\bibitem{GEN-prop}
	H.V. Klapdor-Kleingrothaus et al. 
	MPI-Report MPI-H-V26-1999,
	hep-ph/9910205, in Proc. of the 2nd Int. Conf. on Particle 
	Physics Beyond the Standard Model BEYOND'99, 
	Castle Ringberg, Germany, 6-12 June 1999, 
	eds.  H.V. Klapdor-Kleingrothaus and I.V. Krivosheina, 
	{\it IOP Bristol} (2000) 915-1014.


\bibitem{KK-Neutr98}
	H.V. Klapdor-Kleingrothaus, in Proc. of 18th Int. Conf. on 
	Neutrino Physics and Astrophysics (NEUTRINO 98), 
	Takayama, Japan, 4-9 Jun 1998, (eds) Y. Suzuki et al. 
	{\it Nucl. Phys. Proc. Suppl.} {\bf 77} (1999) 357 - 368.   


\bibitem{HDM97} 
	HEIDELBERG-MOSCOW Coll. (M. G\"unther et al.),  
	{\it Phys. Rev.} {\bf D 55} (1997) 54.


\bibitem{KK-NOW00}
	H.V. Klapdor-Kleingrothaus, 
	{\it Nucl. Phys.} {\bf B 100} (2001) 309-313.

\bibitem{Weinh-Neu00}
	J. Bonn et al., 
	{\it Nucl. Phys.} {\bf B 91} (2001) 273 - 279.



\bibitem{KK-StProc00}
	H.V. Klapdor-Kleingrothaus, 
	in Proc. of the Int. Symposium on Advances in 
	Nuclear Physics, eds.: D. Poenaru and S. Stoica, 
	{\it World Scientific, Singapore } (2000) 123-129.

\bibitem{KK-LP01}
	H.V. Klapdor-Kleingrothaus, 
	{\it Int. J. Mod. Phys.} {\bf A 17} (2002) 3421 -3431, and in 
	Proc. of Intern Conf. LP01, WS 2002, 
	Rome, Italy, July 2001. 



\bibitem{KK-IK-GTF-Valenc03}
	H.V. Klapdor-Kleingrothaus and I.V. Krivosheina, 
	PRHEP-AHEP2003/060, 
	in Proc. of International Workshop on Astroparticle 
	and High Energy Physics, Valencia, Spain, September 2003.

\bibitem{CERN03-GenTF}
	H.V. Klapdor-Kleingrothaus, CERN Courier 43 (2003) p.9.

\bibitem{TF-NIM03}
	H.V. Klapdor-Kleingrothaus, O. Chkvorez, 
	I.V. Krivosheina, H. Strecker, C. Tomei, 
	{\it Nucl. Instr. Meth.} 
	{\bf A 511} (2003) 341 - 346 and hep-ph/0309170, and 
	H.V. Klapdor-Kleingrothaus and I.V. Krivosheina, in Proc.
	of Beyond the Desert 2002, 
	BEYOND02, Oulu, Finland, June 2002, IOP 2003, 
	ed. H.V. Klapdor-Kleingrothaus.  

\bibitem{KK-Modul-NIM03}
	C. Tomei, A. Dietz, I. Krivosheina, H.V. Klapdor-Kleingrothaus, 
	{\it Nucl. Instr. Meth.} 
	{\bf A 508} (2003) 343-352, and hep-ph/0306257.


\bibitem{GenTF-0012022}
	H.V. Klapdor-Kleingrothaus et al., 
	{\it hep-ph/}0103082, {\it NIM} 
	{\bf A 481} (2002) 149-159.

\bibitem{KK-IK}
	H.V. Klapdor-Kleingrothaus and I.V. Krivosheina, in Proc. of 
	``Forum of Physics'', Zacatecas, Mexico, 11-13 May, 2002, 
	eds. D.V. Ahluwalia and M. Kirchbach, 
	{\it Found. Phys.} {\bf 33} (2003) 831-837. 


\bibitem{Trit03}
	C. Weinheimer, in Appec meeting, Karlsrhue, 16-18 September 2003,
	http://www-ik.fzk.de/\%7ekatrin/atw/talks.html, 
	and J. Bonn et al., 
	Nucl. Phys. Proc. Suppl. 110 (2002) 395-397.



\bibitem{Kurch03}
	A.M. Bakalyarov et al. 
	(Moscow grop of HEIDELBERG-MOSCOW ex\-pe\-ri\-ment), 
	hep-ex/0309016.


\bibitem{KK04-Big-Repl}
	H.V. Klapdor-Kleingrothaus et al., in preparation.






\bibitem{DUM-RES-AVIGN-2000}
	C.E. Aalseth et al. (IGEX Collaboration), 
	{\it Yad. Fiz.} {\bf 63, No 7} (2000) 1299 - 1302;
	{\it Phys. Rev.} {\bf D 65} (2002) 092007. 

\bibitem{PRD04-CritIGEX}
	H.V. Klapdor-Kleingrothaus, A. Dietz and I.V. Krivosheina, 
	{\it Phys. Rev.} {\bf D} (2004) and hep-ph/0403056.




\bibitem{KK-NANPino00} 
	H.V. Klapdor-Kleingrothaus, 
	Part. \& Nucl., Lett. 104 (2001) 20-39,  
	\& hep-ph/0102319.



\bibitem{KK-00-NOON-NOW-NANP-Bey97-GEN-prop}
	H.V. Klapdor-Kleingrothaus, hep-ph/0103074 and in 
	Proc. 
	 NOON 2000, 
	World Scientific, Singapore (2001) 219-234.

\bibitem{KK-LowNu2}
	H.V. Klapdor-Kleingrothaus, Proc. 
	LowNu2, 
	World Scientific, Singapore (2001) 116-131 and 
	hep-ph/0104028.

\bibitem{AHLUW02}
	D.V. Ahluwalia in Proc. of Beyond the Desert 2002, 
	BEYOND02, Oulu, Finland, June 2002, 
	IOP 2003, ed. H.V. Klapdor-Kleingrothaus, 143-160;  
	D.V. Ahluwalia, M. Kirchbach, 
	{\it Phys. Lett.} {\bf B 529} (2002) 124.

\bibitem{DAMA-03}
	R. Bernabei et al., 
	Riv. Nuovo  Cim. 26 (2003) 1-73. 



\bibitem{Farj00-04keV}
	D. Fargion et al., 
	in Proc. of DARK2000, Heidelberg, Germany, July 10-15, 
	2000, Ed.  H.V. Klapdor-Kleingrothaus, 
	{\it Springer}, (2001) 455 -468 and 
	in Proc. of Beyond the Desert 2002, BEYOND02, 
 	Oulu, Finland, June 2002, IOP (2003) 543 p., 
	ed. H.V. Klapdor-Kleingrothaus.


\bibitem{MaRaid01}
	E. Ma and M. Raidal,
	{\it Phys. Rev. Lett.} 87 (2001) 011802; 
	Erratum-ibid. 87 (2001) 159901.

\bibitem{Uehara02}
	Y. Uehara, {\it Phys. Lett.} {\bf B 537}  (2002) 256-260 
	and hep-ph/0201277.

\bibitem{Ma-DARK02}
	E. Ma in Proc. of Intern. Conf. 
	on Physics Beyond the Standard Model: Beyond the Desert 02, 
	BEYOND'02, Oulu, Finland, 2-7 Jun. 2002, 
	IOP, Bristol (2003) 95 - 106, ed. H.V. Klapdor-Kleingrothaus.

\bibitem{KAML02}
	KamLAND Coll., 
	{\it Phys. Rev. Lett.} {\bf 90} (2003) 021802 and 
	hep-ex/0212021.

\bibitem{Fogli03}
	G.L. Fogli et al., 
	{\it Phys. Rev.} {\bf D 67} (2003) 073002 and 
	hep-ph/0212127.

\bibitem{BMV02}
	K.S.  Babu, E. Ma and J.W.F. Valle (2002) hep-ph/0206292.

\bibitem{Moh03}
	R.N. Mohapatra, M.K. Parida and G. Rajasekaran, (2003) 
	hep-ph/0301234.

\bibitem{FKR01}
	Z. Fodor, S.D. Katz and A. Ringwald, 
	{\it Phys. Rev. Lett.} {\bf 88} (2002) 171101; 
%
	Z. Fodor et al., {\it JHEP} (2002) 0206:046, or hep-ph/0203198, 
	and in Proc. of Intern. Conf. 
	on Physics Beyond the Standard Model: 
	Beyond the Desert 02, BEYOND'02, Oulu, Finland, 2-7 Jun 2002, 
	IOP, Bristol (2003) 567 p., ed. H V Klapdor-Kleingrothaus and  
	hep-ph/0210123.

\bibitem{CMB02}
	J.E. Ruhl et al., astro-ph/0212229.

\bibitem{WMAP03}
	D.N. Spergel et al., Astrophys. J. Suppl. {\bf 148} (2003) 175 
	and astro-ph/0302209.

\bibitem{Muray03}
	A. Pierce and H. Murayama, hep-ph/0302131. 

\bibitem{Hannes03}
	S. Hannestad, {\it JCAP} {\bf 0305} (2003) 004 and astro-ph/0303076.

\bibitem{Vogel1} 
	P. Vogel in PDG (ed. K Hagiwara et al.) 
	{\it Phys. Rev.} (2002) {\bf D 66} 010001.

\bibitem{Allen03-Wmap}
	S.W. Allen, R.W. Schmidt, S.L. Bridle, 
	{\it Astron. Astrophys.} {\bf 412} (2003) 35-44 and  
	astro-ph/0306386.

\bibitem{S-Sarkar03}
	A. Blanchard, M. Douspis, M. Rowan-Robinson, S. Sarkar, 
	astro-ph/0304237.

\bibitem{Kirch04}
	M. Kirchbach, C. Compean and L. Noriega,  
	hep-ph/0310297, and M. Kirchbach in Proc. of BEYOND03, 
	4th Int. Conf. on Particle 
	Physics Beyond the Standard Model, 
	Castle Ringberg, Germany, 9-14 June 2003,  
	{\it Springer} (2004), ed. H V Klapdor-Kleingrothaus.

\bibitem{XMASS03}
	S. Moriyama (for XMASS collaboration), in Proc. of BEYOND'03, 
	4th Int. Conf. on Particle 
	Physics Beyond the Standard Model, 
	Castle Ringberg, Germany, 9-14 June 2003,  
	{\it Springer} (2004), ed. H V Klapdor-Kleingrothaus.


\bibitem{Hof04}
	R. Hofmann, hep-ph/0401017 v.1.


\bibitem{Barger03}
	M. Tegmark et al., astro-ph/0310723.

\end{thebibliography}
\end{document}